\documentstyle[12pt,epsf]{article}
\newcommand{\bbox}[1]{\mbox{\boldmath$#1$}}
\begin{document}

\begin{center}
{\bf Chiral Metal as a Ferromagnetic Super Spin Chain}\\[7pt]
Leon Balents, Matthew P. A. Fisher and Martin R. Zirnbauer \\[5pt]
Institute for Theoretical Physics, University of California, Santa
Barbara, CA 93106--4030 \\
\date{\today}
\end{center}

\begin{abstract}
The electrons on the surface of a disordered multi-layer integer
quantum Hall system constitute an unusual chiral metal with ballistic
motion transverse to the field, and diffusive motion parallel to it.
We present a non-perturbative analytic treatment of an appropriate
model, consisting of disordered chiral Fermions in two dimensions.  A
supersymmetric generating functional is set up for the correlation
functions of this system.  The strong disorder limit is mapped into a
supersymmetric spin chain, with ferromagnetic exchange coupling,
reflecting the electron's chiral motion.  The ferromagnetic ground
state and the spin wave excitations, corresponding to the diffusion
modes of the chiral metal, are found exactly.  The parametric density
of states correlator in the ergodic limit is computed from a
Boltzmann-weighted sum over low energy spin states.  The result is of
a universal form and coincides with that for a Hermitian random
matrix.
\end{abstract}

{PACS:  73.20.-r, 73.40.Hm, 75.10.Jm}



\section{Introduction}

Disorder has a profound effect on low dimensional electron transport,
generically leading to localization of all states in both one and two
dimensions (2d)\cite{leerama}.  In the quantum Hall regime with a strong
magnetic field, however, the behavior is richer\cite{Prange-Girvin}.
Precisely at the transition between successive integer Hall plateaus,
the 2d electron states are {\it not} localized, but quasi-extended
with multi-fractal scaling characteristics\cite{Bodo}.  Within a Hall
plateau, the 1d edge states are extended, due to their purely chiral
nature\cite{Halperin,Buttiker}.  Indeed, for a single edge mode,
impurities simply lead to an unimportant forward scattering phase
shift.  However, when multiple edge modes are present, such as at the
boundaries of hierarchical fractional quantum Hall effect (FQHE)
states\cite{Wen}, interchannel impurity scattering can be
important\cite{Kane}.  For example, backscattering between the two
counter propagating modes of a $\nu=2/3$ state is predicted to drive
a transition into a phase with charge and neutral sectors
decoupled\cite{KFP}.  Multiple edge modes are also present in
multi-layer systems exhibiting a 3d bulk QHE in perpendicular
applied field.  Together, these edge modes comprise a
conducting two-dimensional subsystem, which has been the focus of
recent attention\cite{Chalker,Balents96}.  Such systems can be
realized by fabricating multi-layer GaAs
heterostructures\cite{Stormer}, but occur naturally in the Bechgaard
salts, a class of quasi 1d compounds which exhibit a cascade of
field-induced spin-density-wave transitions between bulk QHE
phases\cite{SDWexpts}.

The chiral 2d metal ``living" on the surface of such bulk QHE systems
can be probed via transport experiments along the field direction
($z$-axis).  For the integer quantum Hall effect (IQHE), the edge
state in each layer is a free (chiral) Fermion, so that uniform
inter-layer tunnelling leads to coherent $z$-axis motion.  The low
energy surface excitations then comprise a 2d Fermi surface, which is
equivalent to ``half" of a conventional 2d open Fermi surface.  A
standard diagrammatic approach\cite{Balents96} shows that impurity
scattering leads to diffusive motion along the $z$-axis, with
ballistic motion perpendicular to the field.  Since backscattering is
{\it not} possible in the ballistic direction, conventional
localization effects are not expected.  Indeed, a perturbative
expansion about the diffusive metal shows an absence of any
localization\cite{Balents96}.  Moreover, random-walk arguments and
numerics on an appropriate network model\cite{Chalker}\ suggest that
this conclusion is valid generally.

In this paper we describe a non-perturbative analytic treatment of the
IQHE surface sheath with impurity scattering.  We show that an
appropriate generating functional for the 2d electron correlations can
be mapped onto a 1d quantum spin chain.  The replica trick is avoided
by employing Fermionic and Bosonic partners (Bosonic ``ghosts'') from
the outset to perform the desired quenched average.  The resulting
spin chain is thus supersymmetric, involving both Bosonic and
Fermionic spins.  Since the edge modes all move in the same direction,
the super spin chain has a ferromagnetic exchange
interaction\cite{Kim}.  We show that the ground state of this super
spin chain consists of all spins ``aligned'', analogous to the ground
state of an ordinary ferromagnet.  Moreover, there is a class of
one-magnon excitations which are exact eigenstates of the super spin
chain, leading to a sharp one-magnon pole in the spin-spin correlation
function.  This pole corresponds to the diffuson pole, describing the
diffusive $z$-axis motion of electrons in the chiral 2d system.  The
exact pole confirms the complete absence of all localization
corrections for this 2d chiral metallic system.

The use of supersymmetry has a long history in the theory of
disordered electron transport.  Efetov\cite{efetov} reformulated
earlier replica field theories\cite{Schaefer-Wegner}\ to obtain a
supersymmetric non-linear sigma model (NL$\sigma$M) description of
disordered electron motion.  This approach enabled the computation of
several new properties, most notably electron localization in 1d and
random matrix spectral correlations in 0d.  To describe plateau
transitions in the IQHE, a topological term is needed in the
$\sigma$-model, as originally argued by Pruisken
et. al.\cite{pruisken} and incorporated into the supersymmetric
approach by Weidenm\"uller\cite{haw}.  More recently, one of
us\cite{mrz-iqhe} has shown that the 2d supersymmetric $\sigma$-model
with topological term can be mapped into a 1d super spin chain
Hamiltonian, with antiferromagnetic exchange interaction.  This spin
chain was, in turn, shown to be equivalent to the Chalker-Coddington
network model\cite{Chalker-Codd,DHLee}.  The antiferromagnetic
exchange corresponds to counter propagating edge modes of the network
model, in contrast to the ferromagnetic exchange for the chiral 2d
metal studied here.  Unfortunately, the AFM super spin chain model is
{\it much} more complicated than the ferromagnetic model analysed
below, and has so far eluded analysis.

In addition to obtaining the exact ground state and one-magnon
excitations, we employ the ferromagnetic super spin chain to extract
spectral correlations of the chiral metal.  A finite length spin chain
at non-zero temperatures corresponds to a 2d chiral metal with both
dimensions finite.  As with other zero-dimensional random electron
systems, we expect interesting universal spectral
correlations\cite{sla}.  Working directly with the 1d spin
Hamiltonian, we derive expressions for the two-point spectral
correlation function and the parametric correlations.  Not
surprisingly, the results coincide with the universal random matrix
theory results for the unitary ensemble, obtained previously from
other methods\cite{mehta,Simons93}.  Our approach is notable for its
(relative) simplicity, involving a simple sum over a large
(super-)spin multiplet of quantum states.

This paper is organized as follows.  After introducing the model in
Section II, we construct the supersymmetric generating functional in
Section III, and show that in the limit of strong disorder it can be
mapped into a super spin chain, with ferromagnetic exchange.  In
Section IV we consider the electron's motion in the thermodynamic
limit.  From the exact ground state and one-magnon spectrum of the
spin chain, the diffuson (or density-density correlator) can be
computed exactly, revealing diffusive motion parallel to the field,
and ballistic motion perpendicular to it.  In Section V we consider
mesoscopic effects for a sample with finite dimensions.  Due to the
system's chirality and anisotropy there are three mesoscopic regimes.
In the ``zero-dimensional" ergodic regime, we compute the two-point
spectral correlation function and the parametric correlations, by
summing over a zero-energy multiplet of super spin states.  Section VI
is devoted to a summary and brief conclusions.

\section{Model}

\begin{figure}[hbt]
\hspace{2.0in}\epsfxsize=3.0truein\epsfbox{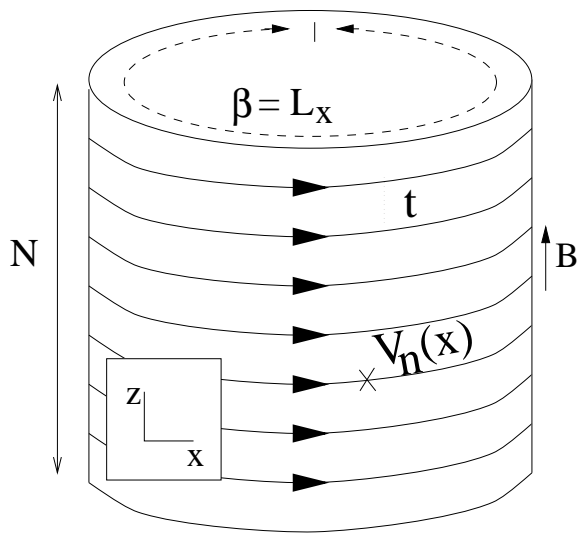}
\vspace{15pt}
\caption{Geometry of a 3d quantum Hall sample.  $z$--axis
transport is included via the tunneling amplitude $t$, and impurity
scattering by the random potential $V$.  The circumference is denoted
$\beta = L_x$, and the height in units of the layer spacing is $N$.}
\label{fig1}
\end{figure}

The system of interest (see Fig.~1) is a collection of $N$
edge states, each described by a chiral Fermion, which are coupled
together by an inter-edge hopping strength, $t$:
\begin{equation}
{\cal H}_0 = \sum_{n = 1}^N \int_0^\beta dx [\psi^\dagger_n i 
\partial_x^{\vphantom{\dagger}} \psi_n^{\vphantom{\dagger}} -
t(\psi^\dagger_n \psi_{n+1}^{\vphantom{\dagger}} + {\rm h.c.})]  ,
\end{equation}
where $\beta = L_x$ is the circumference of the surface sheath (and
the sample).  Our choice of units is such that the edge velocity
$v=1$.  The full Hamiltonian including random scattering is ${\cal H} =
{\cal H}_0 + {\cal H}_1$, where
\begin{equation}
{\cal H}_1 = \sum_n \int dx V_n(x) \psi^\dagger_n
\psi_n^{\vphantom{\dagger}} .
\end{equation}
For simplicity the random potential is assumed to have a Gaussian 
distribution with zero mean and
\begin{equation}
[V_n(x) V_{n^\prime}(x^\prime)]_{\rm ens} = 2u \delta_{nn^\prime}
\delta(x-x^\prime) ,
\end{equation}
where the square brackets denote an ensemble average over disorder
realizations.

\section{Generating functional}

As usual, a generating functional can be introduced from which the
electron Green's functions can be extracted\cite{Schaefer-Wegner}.  Of
interest are products of retarded and advanced Green's functions,
denoted $G_\pm$, where
\begin{eqnarray}
&&G_\pm(n,n';x,x';\pm\omega/2) 
\nonumber \\
&=& \langle n x |\left[{\cal H} \pm i
(\eta+i\omega/2)\right]^{-1}|n' x' \rangle .
\end{eqnarray}
Here $\eta$ is a positive infinitesimal, and $\omega/2$ is the
electron energy.  The appropriate generating functional can be written
as an integral over a spinor of Grassmann fields, $\psi_{n\alpha}$,
where $\alpha = \pm$ for retarded (advanced).  To make the analogy
with spin, we will sometimes use instead the notation $\alpha =
\uparrow , \downarrow$.  To avoid the use of the replica method, we
adopt the supersymmetric approach, and consider an additional
integration over a spinor of complex fields, $\phi_{n\alpha}$.  The
supersymmetric generating functional, ${\cal Z}$, introduced below, is then
normalized, with ${\cal Z}=1$ for every realization of the disorder
potential.  An average over an ensemble of disorder realizations can
then be safely performed.

To insure convergence of the functional integrals over $\phi$, the
contribution to ${\cal Z}$ in the retarded sector is defined by
exponentiating $i({\cal H} -\omega/2+ i\eta)$, whereas in the advanced
sector, $- i({\cal H} +\omega/2- i\eta)$.  The appropriate
supersymmetric generating functional then takes the form
\begin{eqnarray}
{\cal Z} &=& \int D\psi D\overline{\psi} D\phi D\phi^* \exp(-S) ,
\\
{\rm with} \quad S &=& \int dx \sum_n {\cal L}_n .
\end{eqnarray}
The Lagrangian is expressed as a sum of Boson and Fermion
contributions:
\begin{equation}
{\cal L}_n = L(\phi^*,\phi) + L(\overline{\psi},\psi) ,
\end{equation}
with
\begin{eqnarray}
L(\phi^*,\phi) &=& \phi_n^* \sigma_z^{\vphantom{*}} 
(\partial_x^{\vphantom{*}} - iV_n^{\vphantom{*}}) \phi_n^{\vphantom{*}}
\nonumber \\
&+& it(\phi_n^*
\sigma_z^{\vphantom{*}} \phi_{n+1}^{\vphantom{*}} 
+ {\rm c.c.}) + \tilde\eta \phi_n^* \phi_n^{\vphantom{*}},
\label{Lagrangian}
\end{eqnarray}
where $\tilde\eta = \eta+ i\omega/2$.  In the future, we will drop the
tilde, remembering to let $\eta \rightarrow \eta + i\omega/2$ when
necessary in computing quantities at non-zero frequency.  By
construction, ${\cal Z}=1$, independent of the random potential.

The single particle Green's functions can be expressed as averages of
either the complex or Grassmann fields.  Specifically,
\begin{eqnarray}
G_\alpha (n,n^\prime;x,x^\prime) &=& - \alpha i <\phi_{n \alpha}(x)
\phi^*_ {n^\prime \alpha}(x^\prime) > \\
&=& - \alpha i <\psi_{n\alpha}(x)
\overline{\psi}_{n^\prime \alpha}(x^\prime) > ,
\end{eqnarray}
where $\alpha = \pm$ and the brackets denote an average taken with
weight $\exp(-S)$.  The density of states (DOS) follows from the
imaginary part of the diagonal element,
\begin{equation}
\rho = \sum_{n=1}^N {\beta \over 2\pi i} (G_- - G_+)(n,n;x,x).
\label{dos_total}
\end{equation}
Note that an integral over $x$ is unnecessary, since the DOS is
independent of $x$.  This follows from the chiral conservation law,
\begin{equation}
\partial_x |\Phi|^2 =
\nabla_n \cdot J,
\label{continuity_eqn}
\end{equation}
where the current is $J(x,n) = 2t {\rm Im}[\Phi^*(x,n)\Phi(x,n+1)]$.
Eq.~(\ref{continuity_eqn}) is valid for every eigenfunction $\Phi$ of
the Hamiltonian.  To resolve the ambiguity which arises because the
DOS involves $G$ at equal points, $x=x^\prime$, we take the
symmetrized form,
\begin{eqnarray}
4 \pi \rho &=& \beta \sum_{n,\alpha} \sum_\pm <\phi_{n\alpha}(x \pm \epsilon)
\phi^*_{n\alpha}(x) > 
\label{symmetrized} \\
&=& \beta \sum_{n,\alpha} \sum_\pm <\psi_{n\alpha}(x \pm \epsilon)
\overline{\psi}_{n\alpha}(x) > ,
\end{eqnarray}
with $\epsilon$ a positive infinitesimal.

\subsection{Coherent State}

Following D.H. Lee\cite{DHLee}, we reinterpret the spatial coordinate
$x$ as an imaginary time coordinate $\tau$, with $S$ the Euclidean
action for a 1d quantum system.  The Hamiltonian for this 1d quantum
system, denoted $H$, acts as a transfer matrix in the $x$ direction.
In order to extract $H$, it is convenient to recast the action $S$
into the canonical form for a coherent state path integral of a 1d
system of Bosons and Fermions.  Up to a sign in the advanced sector,
the linear $x$-derivative term in (8) is already in the appropriate
form.  To ``correct" this sign, we transform the advanced complex
fields as
\begin{equation}
\phi_{n\downarrow} \leftrightarrow \phi_{n\downarrow}^*  ,
\label{transformation_B}
\end{equation}
leaving the retarded fields, $\phi_{n \uparrow}$, unchanged.  The
first term in the Lagrangian then takes the canonical form: $\phi_n^*
\partial_\tau^{\vphantom{*}} \phi_n^{\vphantom{*}}$.  
In the Fermion sector we similarly transform the Grassmann fields:
\begin{eqnarray}
\psi_{n\downarrow} &\rightarrow& - \overline{\psi}_{n\downarrow}  , 
\\
\overline{\psi}_{n\downarrow} &\rightarrow& \psi_{n\downarrow} ,
\label{transformation_F}
\end{eqnarray}
again leaving the retarded Grassmann fields unchanged.  The transformed
Lagrangian can be written ${\cal L}_n = L_0 + L_t + L_V + L_\eta$ ,
with
\begin{equation}
L_0 = \phi_n^* \partial_\tau^{\vphantom{*}} \phi_n^{\vphantom{*}} 
+ \overline{\psi}_n \partial_\tau^{\vphantom{*}} \psi_n^{\vphantom{*}}
\label{L_0}
\end{equation}
in canonical form.  The term
\begin{equation}
L_\eta = \eta (\phi_n^* \phi_n^{\vphantom{*}} + 
\overline{\psi}_n \psi_n^{\vphantom{*}} ) ,
\label{L_eta}
\end{equation}
and the random potential,
\begin{equation}
L_V = -iV_n(\tau) (\phi_n^* \sigma_z^{\vphantom{*}} \phi_n^{\vphantom{*}} 
+ \overline{\psi}_n \sigma_z^{\vphantom{*}} \psi_n^{\vphantom{*}} ) ,
\end{equation}
are unchanged, whereas 
the tunnelling terms become
\begin{equation}
L_t = it (A_{n+1 , n} + A_{n , n+1})
\end{equation}
with
\begin{equation}
A_{m , n} = \phi_{m\uparrow}^* \phi_{n\uparrow}^{\vphantom{*}} -
\phi^*_{n\downarrow} \phi_{m\downarrow}^{\vphantom{*}} + 
(\phi \rightarrow \psi)  .
\end{equation} 

\subsection{ 1d Hubbard model}

We can now perform an average over the ensemble of random potentials.
This gives
\begin{equation}
L_u = u  \left( \phi_n^* \sigma_z^{\vphantom{*}} \phi_n^{\vphantom{*}} 
+ \overline{\psi}_n \sigma_z^{\vphantom{*}} \psi_n^{\vphantom{*}}
\right)^2 .
\label{L_u}
\end{equation}
The full action $S$ is now equivalent to the Euclidian action for a 1d
quantum system, and the 1d quantum Hamiltonian (transfer matrix) can
be readily extracted.  One simply replaces the Grassmann fields
$\psi_n$ by Fermion operators, $f_n$, and the complex fields,
$\phi_n$, by Bose operators, $b_n$.  The resulting 1d Hamiltonian
takes the form: $H = H_t + H_u + H_\eta$ with
\begin{eqnarray}
H_\eta &=& \eta \sum_n ( b^\dagger_n b_n^{\vphantom{\dagger}} + 
f_n^\dagger f_n^{\vphantom{\dagger}}) ,
\label{Hetabf_def} \\
H_u &=& u \sum_n \left( b_n^\dagger \sigma_z^{\vphantom{\dagger}} 
b_n^{\vphantom{\dagger}} + f_n^\dagger \sigma_z^{\vphantom{\dagger}} 
f_n^{\vphantom{\dagger}} \right)^2 ,
\\
H_t &=& it \sum_n (A_{n+1,n} + A_{n,n+1}) ,
\end{eqnarray}
where
\begin{equation}
A_{n+1,n} = b^\dagger_{n+1\uparrow} b_{n\uparrow}^{\vphantom{\dagger}} 
- b^\dagger_{n\downarrow} b_{n+1\downarrow}^{\vphantom{\dagger}} 
+ (b \rightarrow f) ,
\label{Anbf_def}
\end{equation}
and the Fermion and Boson operators satisfy canonical commutation
relations
\begin{eqnarray}
&&[b_{n\alpha}^{\vphantom{\dagger}}, b_{n^\prime \beta}^\dagger ] 
= \delta_{n n^\prime} \delta_{\alpha\beta} , \\
&&[f_{n\alpha}^{\vphantom{\dagger}}, f_{n^\prime \beta}^\dagger ] 
= \delta_{nn^\prime} \delta_{\alpha\beta} .
\end{eqnarray}
Here, and in the remainder of the paper, $[{\cal O}_1, {\cal O}_2]$
denotes the graded or super- commutator, defined as
\begin{equation}
  [{\cal O}_1,{\cal O}_2] \equiv {\cal O}_1 {\cal O}_2 - (-1)^{|{\cal
      O}_1||{\cal O}_2|} {\cal O}_2{\cal O}_1,
\end{equation}
where $|{\cal O}| = 0$ if ${\cal O}$ is a Bosonic operator, while
$|{\cal O}| = 1$ if it is a Fermionic one.  This 1d Hamiltonian
describes spinful Bosons and Fermions hopping on a 1d lattice, with
hopping strength $t$, interacting via a quartic term with strength
$u$.  Large interaction $u$ corresponds to strong disorder.  For large
disorder $u$, this term constrains the on-site Bose and Fermion
densities to satisfy
\begin{equation}
b^\dagger_\uparrow b_\uparrow^{\vphantom{\dagger}} 
+ f^\dagger_\uparrow f_\uparrow^{\vphantom{\dagger}} =
b^\dagger_\downarrow b_\downarrow^{\vphantom{\dagger}} 
+ f^\dagger_\downarrow f_\downarrow^{\vphantom{\dagger}} , 
\label{Hubbard_constraint}
\end{equation}
on each site $n$.  There is a large energy cost ($u$) for
configurations involving different numbers of particles in the
retarded and advanced sectors, whereas retarded and advanced particles
can move together.  This reflects the phase cancellation between the
retarded and advanced Green's functions.  The positive infinitesimal
restricts the total number of particles.

The similarity between the full Hamiltonian and more familiar
interacting 1d models such as the Hubbard model\cite{Fradkin}\ can be
revealed by introducing a new set of operators.  To this end, we
define
\begin{equation}
B_\uparrow = b_\uparrow , \quad B_\downarrow = b_\downarrow^\dagger ,
\end{equation}
and
\begin{equation}
F_\uparrow = f_\uparrow , \quad F_\downarrow = f_\downarrow^\dagger .
\end{equation}
The $F$ operators are bona fide Fermion operators, satisfying
\begin{equation}
[F_{n\alpha}^{\vphantom{\dagger}}, F_{n^\prime \beta}^\dagger ] 
= \delta_{nn^\prime} \delta_{\alpha\beta} ,
\end{equation}
but the operator $B_\downarrow$ does {\it not} satisfy the canonical
Boson commutator, but rather, $[B_\downarrow^{\vphantom{\dagger}},
B_\downarrow^\dagger] = -1$.  To restore the canonical form we define
\begin{equation}
\overline{B}_\downarrow = - B_\downarrow^\dagger ,
\end{equation}
and also $\overline{B}_\uparrow = B_\uparrow^\dagger$, which now
satisfy
\begin{equation}
[B_{n\alpha}, \overline{B}_{n^\prime \beta} ] = \delta_{nn^\prime}
\delta_{\alpha
\beta}  .
\end{equation}
As defined, $B_n$ and $\overline{B}_n$ satisfy canonical Bose
commutation relations, but it must be kept in mind that
$\overline{B}_n$ is {\it not} the adjoint of $B_n$.

In terms of these new operators, the 1d Hamiltonian takes a form which
closely resembles the Hubbard model, with spin-independent hopping and
interaction:
\begin{eqnarray}
H_t + H_u &=& it \sum_n (A_{n+1,n} + A_{n,n+1}) 
\nonumber \\
&&+ u \sum_n \left( \overline{B}_n B_n + F_n^\dagger 
F_n^{\vphantom{\dagger}} \right)^2 ,
\end{eqnarray}
where
\begin{equation}
A_{n+1,n} = \overline{B}_{n+1} B_n + F_{n+1}^\dagger 
F_n^{\vphantom{\dagger}} .
\label{AnBF_def}
\end{equation}
The positive infinitesimal term now acts like a small magnetic field,
coupling to the $z$-component of the ``spin":
\begin{equation}
H_\eta = \eta( \overline{B}_n \sigma_z B_n 
+ F_n^\dagger \sigma_z^{\vphantom{\dagger}} F_n^{\vphantom{\dagger}})
\label{HetaBF_def}  .
\end{equation}

As with the Hubbard model, one anticipates that the interaction term
will lead to a ``charge gap", strongly suppressing fluctuations in
$\overline{B}_n B_n + F_n^\dagger F_n^{\vphantom{\dagger}}$.  Of
interest are the remaining gapless ``spin" excitations.  For the
conventional Fermionic Hubbard model, these spin excitations can be
revealed by employing a transformation, to map the
Hubbard model into a Heisenberg spin chain\cite{Fradkin}.  A similar transformation
is desirable in this 1d supersymmetric model.  To this end, we will
return to the path integral representation, and perform a gauge
transformation before averaging over disorder.  This will allow us to
obtain directly a (supersymmetric) spin chain model.

\subsection{Gauge transformation}

Consider the gauge transformation
\begin{equation}
\phi_n \rightarrow \exp(i \sigma_z \int_{- \infty}^\tau d\tau^\prime 
V_n(\tau^\prime)) \phi_n ,
\end{equation}
and an identical transformation for $\psi_n$.  Due to the linear
derivatives in $L_0$, this eliminates the on-site random potential
term in $L_V$, at the expense of introducing randomness into the
hopping term.  The full Lagrangian becomes ${\cal L}_n = L_0 + L_t +
L_\eta$ , with $L_0$ and $L_\eta$ as given in Eq.~(\ref{L_0}) and 
Eq.~(\ref{L_eta}), and
\begin{equation}
L_t = i (t_n A_{n+1,n} + t_n^* A_{n,n+1}) .
\end{equation}
The tunnelling amplitudes, $t_n(\tau)$, are both random and
complex, given by
\begin{equation}
t_n(\tau) = t \exp\left( i \int_{- \infty}^\tau d\tau^\prime
[ V_n(\tau^\prime)-V_{n+1}(\tau^\prime) ] \right) .
\end{equation}
They satisfy
\begin{equation}
[t_n(\tau) t^*_{n^\prime} (\tau^\prime)]_{\rm ens} = \delta_{nn^\prime}
t^2 e^{-2u|\tau-\tau^\prime|} .
\end{equation}
In the following we will focus on the limit of strong disorder (large
$u$), and approximate the above exponential with a delta function:
\begin{equation}
[t_n(\tau) t^*_{n^\prime} (\tau^\prime)]_{\rm ens} \rightarrow
D \delta_{nn^\prime} \delta(\tau-\tau^\prime)  ,
\end{equation}
where $D = t^2/u$.  The quantity $D$ will play the role of the exchange
interaction in the supersymmetric spin chain.  The above large $u$
limit is then analogous to passing from a Hubbard model to a
spin chain.

To extract the 1d (spin chain) Hamiltonian, we again perform an
ensemble average over disorder.  Since the above delta function is
really a short-range symmetric function of its argument, we will
arrive at a symmetrized form.  On passing again to Bose and Fermi
operators, the full 1d Hamiltonian is given by $H = H_D + H_\eta$
where $H_\eta$ was defined in Eq.~(\ref{HetaBF_def}), and
\begin{equation}
H_D = {D\over 2} \sum_n (A_{n+1,n} A_{n,n+1} + A_{n,n+1} A_{n+1,n}) ,
\label{HD_def}
\end{equation}
with $A_{n+1,n}$ as before, in Eq.~(\ref{AnBF_def}).

Before recasting the Hamiltonian $H_D$ into the form of a spin chain,
it is instructive to re-express the DOS in terms of the Fermion and
Boson operators.  Recalling Eq.~(\ref{symmetrized}) we obtain
for the density of states, $\rho = \sum_n \rho_n$ with
\begin{equation}
4 \pi \rho_n / \beta = \langle 
b_{n\uparrow}^\dagger b_{n\uparrow}^{\vphantom{\dagger}} 
+ b_{n\uparrow}^{\vphantom{\dagger}} b_{n\uparrow}^\dagger 
+ b_{n\downarrow}^\dagger b_{n\downarrow}^{\vphantom{\dagger}} 
+ b_{n\downarrow}^{\vphantom{\dagger}} b_{n\downarrow}^\dagger 
\rangle ,
\end{equation}
where $\langle ... \rangle$ is a ``thermodynamic'' expectation
value with Hamiltonian $H$, see Sec.~\ref{bc_ss}\ for the details.
A similar expression exists in terms of the Fermion operators
$b \rightarrow f$.  Equivalently we have
\begin{equation}
2 \pi \rho_n / \beta = 1 + \langle b^\dagger_n b_n^{\vphantom{\dagger}} 
\rangle = 1 - \langle f^\dagger_n f_n^{\vphantom{^\dagger}} \rangle ,
\end{equation}
which involves the on-site Boson and Fermion densities.  These
expressions can also be written as
\begin{equation}
2 \pi \rho_n / \beta = \langle \overline{B}_n \sigma_z B_n \rangle = - \langle 
F^\dagger_n \sigma_z^{\vphantom{\dagger}} F_n^{\vphantom{\dagger}} \rangle .
\end{equation}
As we shall now show, these correlators can be naturally interpreted
as the $z$-component of an appropriately defined spin operator.  In
terms of these spin operators, the Hamiltonian $H_D$ will take the
form of a ferromagnetic (super-)spin chain.
 
\subsection{(Super-)spin chain}

To cast $H_D$ in spin chain form, it is useful to introduce
generalized spin currents.  This can be done in a supersymmetric
manner, by defining a four-component superfield,
\begin{equation}
\Psi_n = (F_n, B_n), \; \; \overline{\Psi}_n = 
(F^\dagger_n,\overline{B}_n^{\vphantom{\dagger}}).
\end{equation}
When necessary, we will use latin indices ($a,b,\cdots$) to denote the
Fermion/Boson label, i.e. $a= F,B \leftrightarrow 1,2$.  The $\Psi$
obey the mixed statistics relation
\begin{equation}
\Psi_{m a \alpha}\overline{\Psi}_{n b \beta} = M_{ab} \overline{\Psi}_{n b
\beta} \Psi_{m a \alpha} + \delta_{mn}\delta_{ab}\delta_{\alpha\beta},
\label{commutators}
\end{equation}
where $M_{ab} = \sigma^x_{ab} - \sigma^z_{ab}$.  We may then define a
four-vector super spin matrix via
\begin{equation}
{\cal J}^\mu_{ab} = \overline{\Psi}_a \gamma^\mu \Psi_b =
\overline{\Psi}_{a\alpha}
\gamma^\mu_{\alpha\beta} \Psi_{b\beta},
\end{equation}
where $\gamma = (\bbox{1},\bbox{\sigma})/2$.  In matrix form, this is
\begin{equation}
{\cal J}^\mu = \left(\begin{array}{cc} F^\dagger \gamma^\mu F &
F^\dagger \gamma^\mu B \\
\bar{B}\gamma^\mu F & \bar{B}\gamma^\mu B
\end{array}\right).
\end{equation}
The global current operators
\begin{equation}
{\cal J}^\mu_{\rm TOT} = \sum_n {\cal J}^\mu_n
\end{equation} 
generate symmetries of $H_D$, as can be easily seen by checking $[{\cal 
J}^\mu_{\rm TOT}, A_{n+1,n} ] = 0$ using Eq.~(\ref{commutators}).  Thus
\begin{equation}
[ {\cal J}^\mu_{\rm TOT}, H_D] = 0.
\label{bigcom}
\end{equation}  

It is now straightforward to obtain the spin chain representation.
Noting that the hopping operator $A_{n+1,n} = \overline{\Psi}_{n+1} 
\Psi_n$, we see that substitution in Eq.~(\ref{HD_def}) yields
\[
H_D = D \sum_n \left( \overline{\Psi}_n\Psi_n -
\overline{\Psi}_{na\alpha}\Psi_{nb\beta}\sigma^z_{bc}\overline{\Psi}_{n+1
c\beta} \Psi_{n+1 a \alpha}\right).
\]
The identities $\overline{\Psi}_{n a \alpha}\Psi_{n b \beta} = 
2 {\cal J}^\mu_{ab}\gamma^\mu_{\beta\alpha}$ and ${\rm Tr}(\gamma^\mu 
\gamma^\nu) = \delta^{\mu\nu}/2$ then give 
\begin{equation}
H_D = 2D \sum_n {\rm Tr} \left[ {\cal J}_n^0 - {\cal J}_n^\mu \sigma_z
{\cal J}^\mu_{n+1} \right].
\end{equation}
The minus sign on the spin-spin interaction indicates that this is a
{\sl ferromagnetic} (super-)spin chain.\footnote{Note that the presence
of $\sigma^z$ is equivalent in the second term to the conventional
``supertrace'' used in supersymmetric notation.}
The $\eta$ term indeed acts as an infinitesimal field
\begin{equation}
H_\eta = 2\eta \sum_n {\rm Tr} {\cal J}_n^z.
\label{Hetadef}
\end{equation}
The term $\sum_n {\rm Tr} {\cal J}_n^0 = {\cal J}_{\rm TOT}^0$
vanishes on the Bose-Fermi vacuum, and commutes with $H_D$ and
$H_\eta$.  It is therefore zero on all low energy states coupled
by the Hamiltonian $H_D + H_\eta$, and will be dropped from all
future expressions.

In order to perform calculations, it is useful to write out the
interactions in a more explicit (but not manifestly supersymmetric)
form.  Dividing the terms into three sectors, $H_D = H_F + H_B + H_X$,
one finds
\begin{eqnarray}
H_F &=& - {D\over 2} \sum_n \left( F_n^\dagger F_n^{\vphantom{\dagger}} 
F_{n+1}^\dagger F_{n+1}^{\vphantom{\dagger}} +
4 \bbox{S}_n\cdot\bbox{S}_{n+1}\right),
\\
H_B &=& {D\over 2} \sum_n \left( \overline{B}_n B_n \overline{B}_{n+1}B_{n+1} 
+ 4 \bbox{S}^B_n \cdot \bbox{S}^B_{n+1} \right), 
\\
H_X &=& {D\over 2} \sum_n \bigg( F_n^\dagger B_n^{\vphantom{\dagger}}
\overline{B}_{n+1}^{\vphantom{\dagger}} F_{n+1}^{\vphantom{\dagger}} 
+ 4 \bbox{S}^X_n \cdot \tilde{\bbox{S}}^X_{n+1} 
\nonumber \\
&&\hspace{1cm} - \overline{B}_{n}^{\vphantom{\dagger}} 
F_{n}^{\vphantom{\dagger}} F_{n+1}^\dagger B_{n+1}^{\vphantom{\dagger}}
- 4 \tilde{\bbox{S}}^X_n \cdot {\bbox{S}}^X_{n+1} 
\bigg),
\end{eqnarray}
where
\begin{eqnarray}
\bbox{S}_n = F_n^\dagger {\bbox{\sigma} \over 2} F_n,  \qquad
\bbox{S}_n^B = \overline{B}_n{\bbox{\sigma} \over 2} B_n, \\
\bbox{S}_n^X = F_n^\dagger {\bbox{\sigma} \over 2}B_n, \qquad
\tilde{\bbox{S}}_n^X = \overline{B}_n {\bbox{\sigma} \over 2}F_n. 
\end{eqnarray}
The Fermionic Hamiltonian, $H_F$, is precisely that of an ordinary
spin-$1/2$ ferromagnetic spin chain\cite{Fradkin}.  In particular, the Fermionic
currents obey the SU(2) algebra,
\begin{equation}
[S_m^{\alpha},S_n^{\beta}] = i\epsilon^{\alpha\beta\gamma}S_n^{\gamma}
\delta_{mn}.
\end{equation}
Since $H_F$ is invariant under global spin rotations, we expect that
the total spin operator
\begin{equation}
\bbox{S}_{\rm TOT} \equiv \sum_n \bbox{S}_n
\end{equation}
commutes with $H_F$.  In fact, Eq.~(\ref{bigcom}) implies
\begin{equation}
[\bbox{S}_{\rm TOT},H_F] = 0 = [\bbox{S}_{\rm TOT},H_D] ,
\end{equation}
i.e. the full $H_D$ is SU(2) invariant.  $H_\eta$ of course breaks the
symmetry explicitly.

Although the Bosonic sector described by $H_B$ appears also to be
SU(2) symmetric, it is not so in the usual sense.  Because of the
unusual definition of conjugation ($\overline{B} \neq B^\dagger$), the
currents $\bbox{S}_n^B$ are not Hermitian.  This non-hermiticity means
that, even though there exists a set of currents forming an SU(2) Lie
algebra and commuting with $H_D$, these cannot be used to generate
{\sl unitary} transformations $\exp{i\bbox{n}\cdot\bbox{S}^B}$.  A
Hermitian set can be defined by
\begin{eqnarray}
J_n^x &=& i S^{B y}_n = (b_{n\uparrow}^\dagger
b_{n\downarrow}^\dagger + b_{n\uparrow} b_{n\downarrow})/2 ,
\\ 
J_n^y &=& -i S^{B x}_n = (b_{n\uparrow}^\dagger b_{n\downarrow}^\dagger -
b_{n\uparrow} b_{n\downarrow})/2i .
\\ 
J_n^z & = & S^{B z}_n = (b_{n\uparrow}^\dagger 
b_{n\uparrow}^{\vphantom{\dagger}} + b_{n\downarrow}^{\vphantom{\dagger}}
b_{n\downarrow}^\dagger)/2 .
\end{eqnarray}
These obey instead the SU(1,1) algebra\cite{Lie}
\begin{eqnarray}
[ J_n^x , J_n^y ] & = & -i J_n^z , \\
\left[ J_n^y , J_n^z \right] & = & i J_n^x , \\
\left[ J_n^z , J_n^x \right] & = & i J_n^y , 
\end{eqnarray}
where the minus sign in the first relation distinguishes SU(1,1) from
SU(2).  In terms of these operators,
\begin{eqnarray}
H_B &=& 2D \sum_n \Big( {\textstyle{1 \over 4}} (b^\dagger_n 
\sigma_z^{\vphantom{\dagger}} b_n^{\vphantom{\dagger}} - 1) 
(b^\dagger_{n+1} \sigma_z^{\vphantom{\dagger}} 
b_{n+1}^{\vphantom{\dagger}}-1) \nonumber \\
&&\hspace{1.2cm} + J_n^z J_{n+1}^z - J_n^x
J_{n+1}^x - J_n^y J_{n+1}^y \Big).
\label{HB_SU11}
\end{eqnarray}
The Lie algebra of SU(1,1) is a well-studied semisimple Lie algebra\cite{Lie}.
Quite generally, such Lie algebras have a so-called quadratic Casimir
operator $\Lambda$, which commutes with all elements of the algebra,
and is the analog of $\bbox{S} \cdot \bbox{S}$ for SU(2).  For SU(1,1),
\begin{equation}
\Lambda = (J^z)^2 - (J^x)^2 - (J^y)^2.
\end{equation}
It is straightforward to verify that $[\Lambda,\bbox{J}]=0$, and that
therefore the global generator
\begin{equation}
\bbox{J}_{\rm TOT} = \sum_n \bbox{J}_n
\end{equation}
commutes with $H_B$.  As before, in fact,
\begin{equation}
[\bbox{J}_{\rm TOT},H_D] = 0,
\end{equation}
as can be verified by explicit calculation.

\subsection{Boundary Conditions}
\label{bc_ss}

In the above development, we have implicitly assumed that the
coordinate $x$, which runs along the chiral modes, is infinite.  Since
$x$ has been replaced by imaginary time $\tau$, this corresponds to
zero ``temperature", $\beta = \infty$, for the 1d (super-)spin chain.
Consideration of mesoscopic effects requires a finite system, with
both $N$ and $\beta$ finite.  Since the chiral modes cannot end, when
$\beta$ is finite it is necessary to consider periodic boundary
conditions in the $x$ (or $\tau$) direction, as depicted schematically
in Fig.~1.  The boundary conditions in the direction ($n$)
transverse to the chiral currents, can be taken either open,
corresponding to a cylinder, or periodic, corresponding to a torus.

When the system is periodic in the $x$-direction, care is needed in
defining the boundary conditions on the complex and Grassmann fields.
Specifically, the generating functional, ${\cal Z}$, will only equal one if
the determinant from the Grassmann integration precisely cancels the
inverse determinant from the integration over the complex fields.
This requires that the boundary conditions on $\psi$ and $\phi$ be
identical: $\phi(0) = \phi(\beta)$ and $\psi(0) = \psi(\beta)$.  
However, in the standard imaginary time Grassmann path integral
representation for Fermions, the Grassmann fields satisfy
anti-periodic boundary conditions, $\psi(\beta) = - \psi(0)$.  This 
discrepancy prevents us from making the transition
from the path integral to the quantum Hamiltonian
by simply replacing Grassmann fields by Fermionic operators.  To remedy
this, we perform a global gauge transformation,
\begin{equation}
\psi_n(x) \rightarrow e^{i\pi x/\beta} \psi_n(x) ,
\end{equation}
in the Lagrangian Eq.~(\ref{Lagrangian}), which transforms the boundary
conditions for the Grassmann fields from periodic to anti-periodic,
and puts the Grassmann path integral in the standard form we desire.
Under this transformation the Lagrangian picks up an additional term 
of the form
\begin{equation}
L_{bc} = {{i \pi} \over \beta} \bar\psi_n \sigma_z \psi_n .
\end{equation}
Since the transformation Eqs.(\ref{transformation_B}-\ref{transformation_F}) 
leaves this form unchanged, this corresponds to an additional term in 
the 1d Hamiltonian:
\begin{equation}
H_{bc} = {{i \pi} \over \beta} \sum_n f_n^\dagger 
\sigma_z^{\vphantom{\dagger}} f_n^{\vphantom{\dagger}} .
\end{equation}
Equivalently, what we need to do when computing thermodynamic averages
with the Boltzmann weight $\exp (-\beta H)$, 
is to take instead of the usual trace the {\sl supertrace}:
\begin{equation}
\langle O \rangle = {\rm STr} e^{-\beta H} O = {\rm Tr}(-1)^{N_F}
O e^{-\beta H} ,
\end{equation}
counting the contributions from states containing an odd number of 
Fermions, $N_F$, with a minus sign.  This latter procedure is the
one adopted in the following.  Thus we do {\it not} modify the
Hamiltonian, but will remember to insert a minus sign for all states
with odd Fermion number. 

\section{Diffuson}
\label{sec:diffuson}

In this section, we calculate the chiral diffuson in the thermodynamic
($\beta,N \rightarrow \infty$) limit.  It is defined by
\begin{eqnarray}
K_n(x;\eta) &=& \left[ |G_+(n+n',n';x+x',x';0)|^2 \right]_{\rm ens}
\nonumber \\
& = & \left[ G_+(n,0;x,0;0)G_-(0,n;0,x;0) \right]_{\rm ens} .
\end{eqnarray}
It may be calculated in several ways, depending upon which fields
($\phi$ or $\psi$) are used to generate the Green's functions
$G_\pm$.  In particular,
\begin{eqnarray}
K_n(\tau) & = & \langle {\cal T}_\tau S_n^{-}(\tau) S_0^{+}(0)\rangle
\label{KnF}\\ 
& = & \langle {\cal T}_\tau J_n^-(\tau) J_0^+(0) \rangle
\label{KnD} \\ 
& = & - \langle {\cal T}_\tau A_n^-(\tau) {\tilde{A}}_0^+(0)\rangle 
\\ 
& = & \langle {\cal T}_\tau {\tilde{A}}_n^-(\tau) A_0^+(0)\rangle,
\label{KnX} 
\end{eqnarray}
where
\begin{eqnarray}
S_n^\pm &=& S_n^x \pm i S_n^y, \qquad A_n^{\pm} = S_n^{Xx} \pm i S_n^{Xy}, 
\\
J_n^\pm &=& J_n^x \pm i J_n^y, \qquad  {\tilde{A}}_n^{\pm}
= \tilde{S}_n^{Xx} \pm i \tilde{S}_n^{Xy}.
\end{eqnarray} 
These expressions have a simple interpretation when
expressed in terms of the $b$
and $f$ operators, 
\begin{eqnarray}
S_n^+ &=& f_{n\uparrow}^\dagger f_{n\downarrow}^\dagger, \qquad
A_n^{+} = f_{n\uparrow}^\dagger b_{n\downarrow}^\dagger, 
\label{singlet1} \\ 
J_n^+ &=& b_{n\uparrow}^\dagger b_{n\downarrow}^\dagger, \qquad
{\tilde{A}}_n^{+} = b_{n\uparrow}^\dagger f_{n\downarrow}^\dagger
\label{singlet2} .
\end{eqnarray}
The operators create pairs (for the Fermions, these are singlets) of
particles at a particular site.  Eqs.~(\ref{KnF}-\ref{KnX}) indicate
that the diffuson is obtained by creating such a pair at the origin
and propagating it to site $n$ in time $\tau$.  The motion of a pair
describes the long-range coherent propagation, in the chiral
disordered metal, of one retarded and one advanced particle.

As discussed in Sec.~\ref{bc_ss}, the angular brackets in
Eqs.~(\ref{KnF}-\ref{KnX}) indicate an average calculated by a trace
with the weight $\exp(-\beta H)$.  In the thermodynamic limit, $\beta
\rightarrow \infty$, with $\eta$ non-zero, only the zero energy (ground) 
states survive.  Since the Hamiltonian is a sum of positive
semi-definite operators, each of these operators must annihilate any
prospective ground state.  Satisfying this for the two terms in
$H_\eta$ in Eq.~(\ref{Hetabf_def}) therefore implies that the only
state which contributes to the average is the $b-f$ vacuum
$|0\rangle$, defined by
\begin{equation}
b_n |0\rangle = f_n|0\rangle = 0.
\end{equation}
In the Fermionic sector, this is nothing other than the fully
polarized spin-down ferromagnetic state, so that
\begin{equation}
S^z_{\rm TOT}|0\rangle = -{N \over 2}|0\rangle.
\end{equation}
It follows that (as can be easily verified)
\begin{equation}
S^-_{\rm TOT}|0\rangle = 0,
\end{equation}
i.e. $|0\rangle$ is the ``lowest weight'' state of this SU(2)
(sub-) representation.  Similarly, one has
\begin{eqnarray}
J^-_{\rm TOT}|0\rangle & = & 0, \\ J^z_{\rm TOT}|0\rangle & = & {N
\over 2}|0\rangle,
\end{eqnarray}
so this is also the lowest weight state of the SU(1,1)
(sub-)representation, and it is the SU(1,1) analog of the fully polarized 
state.  As in the case of the usual ferromagnet, this is the
{\sl exact} ground state.

Now consider the diffuson.  From Eq.~(\ref{KnF}), we have
\begin{eqnarray}
K_n(\tau) & = & \langle 0| S^-_n e^{-\tau H} S^+_0 |0\rangle \\ & = &
\int_0^{2\pi} {{dkdk'} \over {(2\pi)^2}} \langle 0 | S_k^- e^{-\tau H}
S^+_{k'} |0\rangle e^{ikn},
\label{KnF_1}
\end{eqnarray}
where the ``single magnon'' operator
\begin{equation}
S_k^\pm \equiv \sum_n S_n^\pm e^{\pm i k n}
\end{equation}
simply creates a superposition of local spin-flips with wavevector
$k$.  To evaluate Eq.~(\ref{KnF_1}), we should study the state
\begin{equation}
|k\rangle = S^+_k |0\rangle.
\label{single_magnon}
\end{equation}
Because no $b$ Bosons have been added to the system, this state is
still annihilated by $H_B$ and $H_X$.  The problem then effectively
reduces to a purely Fermionic one, i.e. the usual spin-$1/2$ SU(2)
spin chain.  As in that case, {\sl the state $|k\rangle$ is an exact
eigenstate}!  By direct computation, one finds
\begin{equation}
H|k\rangle = E_k |k\rangle,
\end{equation}
where the dispersion relation is
\begin{equation}
E_k = 2D (1 - \cos k) + 2\eta \approx 2\eta + Dk^2,
\end{equation}
for $k \ll 1$.  This is just the usual ferromagnetic spin-wave
dispersion.  It is thus straightforward to evaluate Eq.~(\ref{KnF_1}),
to obtain
\begin{equation}
K_n(\tau \rightarrow x) = \theta(x)\int_0^{2\pi} {{dk} \over {2\pi}}
e^{-E_k x + i k n}.
\end{equation}
Performing the Fourier transform on $x$ gives the final (exact) result
\begin{eqnarray}
K(k_x,k_z) &=& {1 \over {2\eta + ik_x + 2D(1-\cos k_z)}} 
\nonumber \\
&\approx& {1 \over {2\eta + ik_x + Dk_z^2}}.
\end{eqnarray} 
We thus see that the ferromagnetic dispersion $E_k \approx Dk^2 +
2\eta$ corresponds to the expected form of the anisotropic diffuson
for the chiral metal\cite{Balents96}.

It is interesting to note that this result must also be obtained from
the other formulations, Eqs.~(\ref{KnD}-\ref{KnX}).  In fact, one
finds that there are three other sets of exact single super spin flip
excitations, created by Fourier transforms of the operators $J_n^+$,
$A_n^{X}$, and $\tilde A_n^{+}$ in Eqs.~(\ref{singlet1}-\ref{singlet2}).  
The reader may amuse him or herself by carrying through this calculation 
explicitly.

\section{Mesoscopic Regime}

The above results describe the propagation of density fluctuations in
the thermodynamic limit.  We now consider the level statistics of a
mesoscopic sample.  A natural probe of these statistics is the density
of states correlation function\cite{Simons93},
\begin{equation}
C(\omega;\beta,N) = \left[ \rho(E+\omega/2)\rho(E-\omega/2)
\right]_{\rm ens} - \overline{\rho}^2 .
\label{dos_correlator}
\end{equation}
Of interest is the structure of this correlation function
when the frequency $\omega$ is on the scale of the mean level spacing,
\begin{equation}
\Delta = {{2\pi} \over {N\beta}} ,
\end{equation}
where $\beta v$ is the system size in the $x$-direction, and $v$ is
the edge mode velocity.  For isotropic metallic samples, $C(\omega)$
is a universal function of the ratio $\omega/\Delta$, described by
random matrix theory\cite{mehta,Simons93}.  In the present case, the
sample is both chiral and anisotropic, so care is needed to specify
the random matrix theory regime.  To this end it is useful to consider
several important length scales.

\subsection{Length scales}

For finite $\beta$, the electron takes a finite time, $\tau_B \sim
\beta$, to ballistically traverse the sample in the $x$-direction.
Equating this with the time taken to diffuse across the system
in the transverse direction, $\tau_D \sim N^2/D$,
gives a (transverse) crossover length 
\begin{equation}
L_0 = \sqrt{ D \beta},
\end{equation}
measured in units of the $z$-axis lattice constant.  This length scale
has a simple interpretation in terms of the spin chain.  For a chain
with $N$ sites, the energy of the lowest lying spin wave excitation,
which has wavevector $k_{min} =2\pi/N$, is $E_{min} \sim D (k_{min}^2)
\sim \tau_D^{-1}$.  This energy equals the spin chain ``temperature",
$E_{min} \sim T = 1/\beta$, for a spin chain of length $L_0$.  The
boundary $N = L_0$ thus demarkates the two following regimes: For a
short spin chain with $N < L_0$, finite wavevector spin-wave
excitations will not be appreciably excited, and the system will be
effectively ``zero-dimensional".  The opposite limit, $N > L_0$, which
we refer to as a ``one-dimensional diffusive" regime, corresponds to
the ``classical limit" of the spin chain.  The transverse electron
motion is diffusive.

But there is another important length scale\cite{Chalker}.
A 1d spin chain at finite temperature will have a finite
spin-correlation length, 
\begin{equation}
  \xi \sim D\beta , 
\end{equation}
with $D$ the spin-exchange constant.  This length corresponds to the
electron localization length along the $z$-axis.  In the $\beta
\rightarrow \infty$ limit of interest, one has $\xi \gg L_0$.  The
above ``1d diffusive regime" requires $\xi > N > L_0$.  When $N > \xi$
the system enters into a third regime - a 1d localized regime.

Of the three regimes, the ``zero-dimensional" is the simplest and the
one we focus on.  This regime corresponds to random matrix theory, as
verified below.  Since we consider frequencies on the scale of
$\Delta$, the Heisenberg time $\omega^{-1} \sim \Delta^{-1} \sim
N\beta$.  In this regime, the ergodic condition that $\omega^{-1}$
exceeds both $\tau_B \sim \beta$ and $\tau_D$ requires simply $N \gg
1$.  In the ``1d diffusive regime", corrections to random matrix
theory are anticipated, with wavefunctions possibly exhibiting
multi-fractal scaling behavior, as in other low-dimensional diffusive
systems\cite{Bodo,Falko-Efetov}.

To evaluate the spectral correlator $C(\omega)$ at finite $\beta$
requires performing a full thermodynamic trace over states of the
supersymmetric spin chain -- a complicated task.  Fortunately, in the
``zero-dimensional" regime, all finite wavevector spin-wave
excitations are exponentially suppressed (by $e^{-\beta E_{min}} =
e^{-(2\pi L_0/N)^2}$), and can be ignored.  The trace can thus be
restricted to a subset of states which are annihilated by the
Hamiltonian $H_D$.  This restricted trace is the operator analog of
the zero-mode integration carried out on the conventional
NL$\sigma$M\cite{efetov}.

\subsection{Zero energy multiplet}
\label{sec:zero_energy}

Consider then the {\sl zero energy multiplet} of states
that are annihilated by $H_D$, i.e. the set of ground states in the
absence of the symmetry-breaking field.  This set has a large
degeneracy, because any particular vacuum (e.g. $|0\rangle$) may be
``rotated'' using the symmetry operators ${\cal J}^\mu_{\rm TOT}$ to
obtain another zero energy state.  These operators and the resulting
multiplet form a representation of the ${\rm U}(2|1,1)$
(super)symmetry group of $H_D$, which contains SU(2) and SU(1,1)
sub-representations generated using the operators $\bbox{S}_{\rm TOT}$
and $\bbox{J}_{\rm TOT}$.

To describe the $E=0$ multiplet, it is convenient to use the $b$ Boson
and $f$ Fermion operators, for which $|0\rangle$ is the vacuum.  Of the
16 elements of ${\cal J}$, 4 may be chosen diagonal.  A natural choice
is ${\cal J}_{11}^0$, ${\cal J}_{11}^3$, ${\cal J}_{22}^0$, ${\cal
J}_{22}^3$, or, equivalently, the set of number operators,
\begin{equation}
n_{f\alpha} = \sum_n f_{n\alpha}^\dagger f_{n\alpha}^{\vphantom{\dagger}}, 
\qquad 
n_{b\alpha} = \sum_n b_{n\alpha}^\dagger b_{n\alpha}^{\vphantom{\dagger}}.
\end{equation}
All elements of ${\cal J}$ commute with the total $b - f$ spin operator, 
which is
\begin{equation}
{\rm Tr}{\cal J}^0_{\rm TOT} = {\textstyle{1 \over 2}} \left( n_{f\uparrow} 
- n_{f\downarrow} + n_{b\uparrow} - n_{b\downarrow}\right) = 0,
\end{equation}
where the final equality follows from evaluation in the $|0\rangle$
state.  More fundamentally, this results from the Hubbard constraint
Eq.~(\ref{Hubbard_constraint}).  

The remaining 12 elements of ${\cal J}$ act as raising and lowering
operators.  From Eq.~(\ref{bigcom}), we may act with any string of
these operators upon $|0\rangle$ to produce another zero energy state.
The 12 operators, however, are strongly redundant, and in fact only 4
suffice to generate the full $E = 0$ multiplet.

The first two of these are simply the raising operators in the
SU(2) and SU(1,1) subalgebras, which create delocalized ``spinless''
pairs of Fermions and Bosons, respectively, i.e.
\begin{eqnarray}
S^+_{\rm TOT} & = & \sum_n
f^\dagger_{n\uparrow}f^\dagger_{n\downarrow}, \\ J^+_{\rm TOT} & = &
\sum_n b^\dagger_{n\uparrow}b^\dagger_{n\downarrow}.
\end{eqnarray}
These are just the $k=0$ magnon operators of the previous section.  
In addition, we must consider the Fermionic generators,
\begin{eqnarray}
A_{\rm TOT}^+ &=& \sum_n f_{n\uparrow}^\dagger b_{n\downarrow}^\dagger, \\
{\tilde{A}}_{\rm TOT}^+ &=& \sum_n b_{n\uparrow}^\dagger
f_{n\downarrow}^\dagger.
\end{eqnarray}
These create Fermionic $k=0$ pairs.
To simplify the notation we now drop the bulky subscript TOT. 

To see that $S^+$, $J^+$, $A^+$, and ${\tilde A}^+$, are sufficient to
generate the full $E = 0$ multiplet, note first that the remaining 8
off-diagonal generators all annihilate the vacuum (because they
contain either $b$ or $f$).  Therefore, in any string of generators,
they may be commuted or anticommuted to the right until they
annihilate the vacuum, and it is straightforward to see that in the
process all that may be left is a string of the four operators above.
This proves the completeness of this set.

Furthermore, in any string of $S^+$, $J^+$, $A^+$, and ${\tilde A}^+$, 
the order is unimportant up to the overall sign.  This is because the
commutators
\begin{eqnarray}
[S^+ , J^+] & = & [S^+ , A^+] = [S^+ ,{\tilde A}^+] 
\nonumber \\ 
= [J^+ , A^+] & = & [J^+ , \tilde A^+] = [A^+,{\tilde A}^+] = 0
\end{eqnarray}
all vanish.  A complete basis for the $E = 0$ multiplet is thus formed 
by the states
\[
|n_F^{\vphantom{*}} n_B^{\vphantom{*}} n_A^{\vphantom{*}} 
n_{\tilde A} \rangle = \left(S^+\right)^{n_F}
\left(J^+\right)^{n_B} \left(A^+\right)^{n_A} 
\left({\tilde A}^+\right)^{n_{\tilde A}} |0\rangle.
\]
These states are non-vanishing only if $n_A$ and $n_{\tilde
A}$ are zero or one, as follows from the nilpotency of $A^+$ and
${\tilde A}^+$, e.g.
\begin{eqnarray}
\left(A^+\right)^2 & = & \sum_{m,n} f_{m\uparrow}^\dagger 
b_{m\downarrow}^\dagger
f_{n\uparrow}^\dagger b_{n\downarrow}^\dagger \nonumber \\ & = & -
\sum_{m,n} f_{n\uparrow}^\dagger b_{n\downarrow}^\dagger
f_{m\uparrow}^\dagger b_{m\downarrow}^\dagger = - \left(A^+\right)^2 = 0.
\end{eqnarray}
Likewise, $\big({\tilde{A}}^+\big)^2 = 0$.  
It is thus natural to break up the states into four ladders,
\begin{eqnarray}
&&\left(S^+\right)^{n_F}
\left(J^+\right)^{n_B} |0\rangle, \label{sector0} \\
&&\left(S^+\right)^{n_F}
\left(J^+\right)^{n_B} A^+ |0\rangle, \label{sectorA}\\
&&\left(S^+\right)^{n_F}
\left(J^+\right)^{n_B} {\tilde A}^+|0\rangle, \label{sectorAt}\\
&&\left(S^+\right)^{n_F-1} 
\left(J^+\right)^{n_B-1} A^+ {\tilde A}^+|0\rangle . \label{sectorAAt}
\end{eqnarray}
The allowed values of $n_F$ and $n_B$ in each ladder are essentially
determined by the exclusion principle.  In the first sector,
Eq.~(\ref{sector0}), one may add up to $2N$ Fermions to fill all $N$ sites
with both spin species, so that $n_F = 0,\ldots,N$.  Since any number
of Bosons may live at a site, the sum on $n_B = 0,\ldots,\infty$ is
unbounded above.  This result agrees with what we know from group
theory: these states are simply direct product states of SU(2) and
SU(1,1) representations.  By computing
\begin{equation}
S^z |0\rangle = {1 \over 2}\sum_n \left(f_{n\uparrow}^\dagger
f_{n\uparrow}^{\vphantom{\dagger}} - f_{n\downarrow}^{\vphantom{\dagger}}
f_{n\downarrow}^\dagger\right)|0\rangle = -{N \over 2} |0\rangle,
\end{equation}
and recalling $S^-|0\rangle = 0$, we recognize that $|0\rangle$ is
the lowest weight state of a spin $s = N/2$ SU(2) ladder, which
has the well-known degeneracy $2s+1 = N+1$.  This agrees with the
assignment $S^z(n_F) = -N/2 + n_F$.  Likewise, 
$J^-|0\rangle = 0$, $J^z|0\rangle = {\textstyle{N\over 2}}|0\rangle$,
and $J^z(n_B) = N/2 + n_B$.  Thus $|0\rangle$ can also be regarded
as the lowest weight state of a representation of SU(1,1), with spin
$s = N/2$.  This representation is irreducible, unitary and 
infinite-dimensional, where the third property is necessitated by
the second one, since SU(1,1) is non-compact.

Application of the Pauli principle determines the allowed values of
$n_F$ in the other cases.  In the second and third sectors,
Eqs.~(\ref{sectorA}-\ref{sectorAt}), one Fermion is already present, so
at most $2N-1$ may be added.  Since they come pairwise, we must
restrict $n_F = 0,\ldots,N-1$ in these cases.  Finally, in the fourth
sector, Eq.~(\ref{sectorAAt}), $2N-2$ Fermions may be added,
suggesting that we need only $n_F \leq N$.  However, if $n_F = N$, there
are $N$ Fermionic pairs present, and the Fermionic content of the state 
is uniquely determined.  That is, the full state must be a direct product
of a Bosonic part with the fully occupied Fermionic state.  This
implies that
\[
\left(S^+\right)^N \left(J^+\right)^{n_B} |0\rangle \propto 
\left(S^+\right)^{N-1} \left(J^+\right)^{n_B-1} A^+ {\tilde A}^+ |0\rangle .
\]
We therefore choose $n_F = 1,\ldots,N-1$ in the last sector.

\subsection{Supersymmetric quadruplets}

To calculate traces over the $E = 0$ multiplet, it is beneficial
to organize the ladders of states in a slightly different way.  To
that end, consider the Fermionic operators
\begin{eqnarray}
Q_\alpha &=& \sum_n b_{n\alpha}^\dagger f_{n\alpha}^{\vphantom{\dagger}}, \\
\overline{Q}_\alpha &=& 
\sum_n f_{n\alpha}^\dagger b_{n\alpha}^{\vphantom{\dagger}} ,
\end{eqnarray}
which are particular combinations of the global currents ${\cal
J}_{\rm TOT}^\mu$, and belong to the set of twelve raising and
lowering operators of which we have utilized only $S^+$, $J^+$, $A^+$,
and ${\tilde A}^+$ so far.  The set $\{ Q_\alpha , \overline{Q}_\alpha
, n_{f\alpha} , n_{b\alpha} \}$ where $\alpha = \uparrow$ or $\alpha =
\downarrow$, closes under the graded commutator (i.e. the anticommutator 
for two Fermions, and the commutator in the other cases), and thus
forms a subalgebra of the full algebra of global currents.  This
subalgebra is denoted ${\cal H} = {\rm u}(1|1) \oplus {\rm u}(1|1)$.
(The direct sum is over spin.)  The significance of ${\cal H}$ derives
from the fact that its generators commute not only with $H_D$ but also
with $H_\eta$.  Thus $H_D + H_\eta$ is a Casimir invariant for ${\cal
H}$.  Schur's lemma then states that $H_D + H_\eta$ is a multiple of
unity on every irreducible multiplet of ${\cal H}$.  Let us therefore
construct these multiplets.

We interpret $Q_\uparrow$, $Q_\downarrow$ as lowering operators,
and $\overline{Q}_\uparrow$, $\overline{Q}_\downarrow$ as raising
operators for ${\cal H}$.  A lowest weight state, $|{\rm LW}\rangle$,
for ${\cal H}$ obeys
\begin{equation}
Q_\uparrow | {\rm LW} \rangle = Q_\downarrow | {\rm LW} \rangle = 0 .
\end{equation}
To solve these equations for $| {\rm LW} \rangle$, we take note of
the commutation relations
\begin{eqnarray}
\big[ Q_\uparrow , S^+ \big] &=& \tilde A^+ , \qquad
\big[ \overline{Q}_\uparrow , {\tilde A}^+ \big] = S^+ ,
\label{comrel1} \\ 
\big[ \overline{Q}_\downarrow , J^+ \big] &=& {\tilde A}^+ , \qquad
\big[ {Q}_\downarrow , {\tilde A}^+ \big] = J^+ ,
\label{comrel2} \\ 
\big[ \overline{Q}_\uparrow , J^+ \big] &=& A^+ , \qquad
\big[ {Q}_\uparrow , A^+ \big] = J^+ ,
\label{comrel3} \\ 
\big[ Q_\downarrow , S^+ \big] &=& - A^+ , \quad
\big[ \overline{Q}_\downarrow , A^+ \big] = - S^+ .
\label{comrel4}
\end{eqnarray}
All other graded commutators between the two sets $\{ S^+ , J^+ , A^+ ,
{\tilde A}^+ \}$ and $\{ Q_\uparrow , Q_\downarrow , \overline{Q}_\uparrow
, \overline{Q}_\downarrow \}$ vanish.  Using the above commutation
relations and making the ansatz
\begin{eqnarray}
| {\rm LW} \rangle &=& \alpha 
\left(S^+\right)^{n_1} \left(J^+\right)^{n_2} | 0\rangle 
\nonumber \\
&+& \beta \left(S^+\right)^{n_1-1} \left(J^+\right)^{n_2-1} 
A^+ {\tilde A}^+ | 0\rangle ,
\end{eqnarray}
one finds that the lowest weight condition can be satisfied with
$\alpha = 1$ and $\beta = - n_1$.  We denote the resulting lowest
weight state by $| n_1^{\vphantom{*}} n_2^{\vphantom{*}} , 00 \rangle$.  

Application of the raising operator $\overline{Q}_\uparrow$ to
$| n_1^{\vphantom{*}} n_2^{\vphantom{*}} , 00 \rangle$ produces
\begin{eqnarray}
| n_1^{\vphantom{*}} n_2^{\vphantom{*}} , 
10 \rangle &=& \left(S^+\right)^{n_1} \left(J^+\right)^{n_2-1}
A^+ | 0 \rangle \nonumber \\
&&\propto \overline{Q}_\uparrow | n_1^{\vphantom{*}} n_2^{\vphantom{*}} , 
00 \rangle ,
\end{eqnarray}
while application of $\overline{Q}_\downarrow$ gives
\begin{eqnarray}
| n_1^{\vphantom{*}} n_2^{\vphantom{*}} , 
01 \rangle &=& \left(S^+\right)^{n_1} \left(J^+\right)^{n_2-1}
{\tilde A}^+ | 0 \rangle \nonumber \\
&&\propto \overline{Q}_\downarrow | n_1^{\vphantom{*}} n_2^{\vphantom{*}} 
, 00 \rangle .
\end{eqnarray}
When acting with the raising operator $\overline{Q}_\uparrow$ on 
$| n_1^{\vphantom{*}} n_2^{\vphantom{*}} , 10 \rangle$, or with
$\overline{Q}_\downarrow$ on $| n_1^{\vphantom{*}}
n_2^{\vphantom{*}} , 01 \rangle$, we get zero from
$\overline{Q}_\uparrow^2 = 0 = \overline{Q}_\downarrow^2$ due to the 
Fermionic nature of these operators.
A non-vanishing state results from applying $\overline{Q}_\downarrow$
to $|n_1^{\vphantom{*}} n_2^{\vphantom{*}} , 10 \rangle$:
\begin{eqnarray}
| n_1^{\vphantom{*}} n_2^{\vphantom{*}} , 11 \rangle &=&
\left(S^+\right)^{n_1+1} \left(J^+\right)^{n_2-1} | 0 \rangle
\nonumber \\
&&+ (n_2-1) \left(S^+\right)^{n_1} \left(J^+\right)^{n_2-2} 
A^+ {\tilde A}^+ | 0 \rangle
\nonumber \\
&&\propto \overline{Q}_\downarrow | n_1^{\vphantom{*}} n_2^{\vphantom{*}}
, 10 \rangle .
\label{quadruplet_11}
\end{eqnarray}
Exactly the same state is obtained by applying $\overline{Q}_\uparrow$
to $| n_1^{\vphantom{*}} n_2^{\vphantom{*}} , 01 \rangle$.  The state
Eq.~(\ref{quadruplet_11}) is annihilated by the action of both
$\overline{Q}_\uparrow$ and $\overline{Q}_\downarrow$, which again
follows from $\overline{Q}_\uparrow^2 = 0 =
\overline{Q}_\downarrow^2$.  Thus $| n_1^{\vphantom{*}}
n_2^{\vphantom{*}} , 11 \rangle$ is a highest weight state for ${\cal
H}$.

\begin{figure}[hbt]
\hspace{2.0in}\epsfxsize=3.0truein\epsfbox{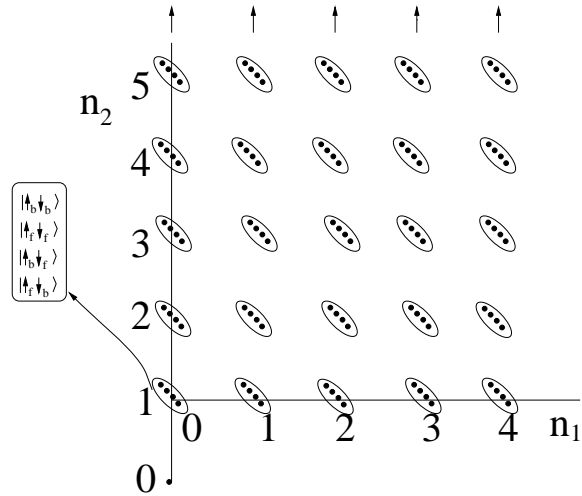}
\vspace{15pt}
\caption{Illustration of the organization of states into
${\cal H}$-quadruplets, shown here for the case $N=5$.  Each
quadruplet contains a state with $n_1$ Fermion pairs and $n_2$
Boson pairs, plus three partners with equal ``spin'' but differing
Boson/Fermion content.  The $n_1=0$, $n_2=1$ quadruplet is shown as an
example.  The vacuum state, with $n_1=n_2=0$ is a unique ${\cal
H}$-singlet.}
\label{fig2}
\end{figure}

The states $| n_1^{\vphantom{*}} n_2^{\vphantom{*}} , \mu\nu \rangle$
$(\mu , \nu = 0, 1)$ so constructed are seen to be in one-to-one
correspondence with the previous set Eqs.~(\ref{sector0}-\ref{sectorAAt}).
What we have thus achieved is an organization of the zero energy
sector into {\sl irreducible ${\cal H}$-quadruplets} (see
Fig.~2), with the exception of only the vacuum $| 0 \rangle$,
which does not fit into the above scheme but figures as a separate
${\cal H}$-singlet.  Each quadruplet consists of two Bosonic $(\mu\nu
= 00 , 11$) and two Fermionic ($\mu\nu = 01 , 10$) states, reflecting
the supersymmetry of the formalism.  Recalling the earlier discussion
based on the exclusion principle, we see that the allowed quantum
numbers for the quadruplets are $n_1 = 0,1,\ldots,N-1$ and $n_2 =
1,\ldots,\infty$.  The pair of quantum numbers $n_1 = n_2 = 0$
corresponds to the ${\cal H}$-singlet vacuum state $|0\rangle$.

By construction, the quadruplet states $| n_1^{\vphantom{*}}
n_2^{\vphantom{*}}, \mu\nu \rangle$ taken together with the vacuum,
constitute an orthogonal basis of the zero energy sector.  As it
stands, they are not normalized to unity.  While it is not difficult
to include the correct normalization factors, there is no need to do
that here.  For our purposes, all we require is that we be able to
calculate traces, and for that linear independence of the states is
entirely sufficient.

\subsection{DOS Correlation Function}

We are finally in a position to calculate correlation functions.  
As a check, consider first the partition function within the zero
energy multiplet.  Since the full partition function equals unity
for all values of the parameter $d$ and approaches the zero mode 
contribution as $d\rightarrow\infty$, the zero mode contribution must 
itself be exactly normalized to unity.  And, indeed, 
\begin{equation}
{\cal Z}_0 = {\rm Tr}_0 (-1)^{N_F} e^{-\beta H} = 1 ,
\end{equation}
where the subscript 0 signifies the restriction to the zero energy
sector.  The 1 on the right-hand side stems from the vacuum.  All
other contributions cancel as $H$ is constant on each ${\cal
H}$-quadruplet, and the two Fermionic states of a quadruplet
come with a minus sign relative to the two Bosonic ones, cf.
Sec.~\ref{bc_ss}.  Since $H_\eta$, by Eq.~(\ref{Hetabf_def}), just
counts the total number of $b$ Bosons and $f$ Fermions, its action
on the quadruplet states is
\begin{equation}
H_\eta | n_1^{\vphantom{*}} n_2^{\vphantom{*}} , \mu\nu \rangle =
2\eta ( n_1 + n_2 ) | n_1^{\vphantom{*}} n_2^{\vphantom{*}} , 
\mu\nu \rangle .
\end{equation}

Now consider the DOS correlator, Eq.~(\ref{dos_correlator}).  Defining
\begin{equation}
G_{\pm n}(\omega) = G_\pm(n,n;x,x;\omega)
\end{equation}
and recalling Eq.~(\ref{dos_total}), we get
\begin{equation}
\rho(\omega/2) = {\beta \over 2\pi i}\sum_n \big[ G_{-n}(\omega/2) 
- G_{+n}(\omega/2) \big] .
\end{equation}

The product therefore becomes
\begin{eqnarray}
&&4\pi^2 \rho(\omega/2)\rho(-\omega/2) / \beta^2 =
\nonumber \\
&&\sum_{nn'} \bigg[ 
G_{+n}(\omega/2)G_{-n'}(-\omega/2) +
G_{-n}(\omega/2)G_{+n'}(-\omega/2) \nonumber \\
&& - G_{+n}(\omega/2)G_{+n'}(-\omega/2)
- G_{-n}(\omega/2)G_{-n'}(-\omega/2) \bigg].
\end{eqnarray}
To proceed, we perform an ensemble average, which may be split into
connected (cumulant) and disconnected pieces.  To calculate the
disconnected terms, we use the symmetric operator representation
\begin{eqnarray}
G_{+n} & = & -{i \over 2}\left(f_{n\uparrow}^{\vphantom{\dagger}}
f_{n\uparrow}^\dagger - f_{n\uparrow}^\dagger
f_{n\uparrow}^{\vphantom{\dagger}}\right) = i f_{n\uparrow}^\dagger
f_{n\uparrow}^{\vphantom{\dagger}} - i/2, \\ G_{-n} & = & -{i \over
2}\left(f_{n\downarrow}^\dagger f_{n\downarrow}^{\vphantom{\dagger}}
-f_{n\downarrow}^{\vphantom{\dagger}} f_{n\downarrow}^\dagger
\right)  = - i  f_{n\downarrow}^\dagger
f_{n\downarrow}^{\vphantom{\dagger}} + i/2.
\end{eqnarray} 
Therefore:
\begin{eqnarray}
\overline{G}_{+n}(\omega/2)\overline{G}_{-n'}(-\omega/2) & + &
\overline{G}_{-n}(\omega/2)\overline{G}_{+n'}(-\omega/2)  \nonumber \\
- \overline{G}_{+n}(\omega/2)\overline{G}_{+n'}(-\omega/2) & - & 
\overline{G}_{-n}(\omega/2)\overline{G}_{-n'}(-\omega/2) = 1,
\end{eqnarray}
where $\overline{G}_{\pm n} \equiv [G_{\pm n}]_{\rm ens} = \mp i/2$.
These disconnected terms thus cancel the $-\overline\rho^2$ term in
the definition of $C(\omega)$, Eq.~(\ref{dos_correlator}).  Then
\begin{eqnarray}
C(\omega) & = & {\beta^2 \over {4\pi^2}} \sum_{n,n'=1}^N 
\bigg[ G_{+n}(\omega/2)G_{-n'}(-\omega/2) +
G_{-n}(\omega/2)G_{+n'}(-\omega/2) \nonumber \\
& & - G_{+n}(\omega/2)G_{+n'}(-\omega/2)
- G_{-n}(\omega/2)G_{-n'}(-\omega/2) \bigg]_{{\rm ens}, c},
\end{eqnarray}
where the subscript $c$ indicates the connected average.

It is well known that the last two terms do not contribute to the
correlator.\footnote{If they are calculated directly using a modified
supersymmetric functional, the resulting Hamiltonian has a unique
ground state even for $\eta=0$, indeed leading to this conclusion.}\
Using this simplification, we arrive at the form
\begin{equation}
C(\omega) = {\beta^2 \over {2\pi^2}} {\rm Re} \sum_{n,n'=1}^N 
\left[ G_{+n}(\omega/2)G_{-n'}(-\omega/2) \right]_{{\rm ens}, c}.
\end{equation}

As with the diffuson, this may be calculated in several ways,
equivalent by supersymmetry.  We arbitrarily choose the Fermion
representation 
\begin{eqnarray}
C(\omega) &=& {{\beta^2}\over {2\pi^2}} {\rm Re}\sum_{nn'} \langle
f_{n\uparrow}^\dagger f_{n\uparrow}^{\vphantom{\dagger}} 
f_{n'\downarrow}^\dagger f_{n'\downarrow}^{\vphantom{\dagger}} \rangle 
\nonumber \\ 
&=& {\beta^2 \over {2\pi^2}}
{\rm Re}\langle n_{f\uparrow} n_{f\downarrow} \rangle.
\label{CFermion0}
\end{eqnarray}
The quadruplet states 
$| n_1^{\vphantom{*}} n_2^{\vphantom{*}} , \mu\nu \rangle$
are eigenstates of $n_{f\uparrow} n_{f\downarrow}$ with eigenvalues
$n_1^2$, $n_1(n_1+1)$, $(n_1+1)n_1$, and $(n_1+1)^2$ for
$\mu\nu = 00$, 01, 10, and 11 respectively.  Therefore the 
trace of $n_{f\uparrow} n_{f\downarrow}$ over the quadruplet
with quantum numbers $n_1 , n_2$ is
\[
n_1^2 - 2n_1 (n_1 + 1) + (n_1 + 1)^2 = 1 .
\]
Using this to evaluate the expectation value Eq.~(\ref{CFermion0}) as a 
trace against the Boltzmann weight, we obtain
\begin{equation}
C(\omega) = {\beta^2 \over {2\pi^2}} {\rm Re}\sum_{n_2=1}^\infty 
\sum_{n_1=0}^{N-1} e^{-\beta(2\eta+i\omega)(n_2+n_1)} .
\label{Cf2}
\end{equation}
While this sum may, of course, be evaluated exactly, it is simpler to
take the continuum limit, valid for $\beta\omega \ll 1$.  Measuring 
energy in units of the level spacing $\Delta= 2\pi/N\beta$, we define
$\tilde\omega = N\beta\omega/2\pi$, and rescaled integration
variables $x=n_2/N$, $y=n_1/N$.  Equation Eq.~(\ref{Cf2}) then becomes
\begin{equation}
C(\omega) = {N^2 \beta^2 \over {2\pi^2}} {\rm Re} \int_0^\infty \! dx
\int_0^1 \! dy \ e^{-2\pi i \tilde{\omega} (x+y)} .
\end{equation}
After division by $\bar\rho^2 = \Delta^{-2}$, and omission of a 
$\delta(\tilde\omega)$ term, this result takes the universal form
\begin{equation}
C(\omega) / \bar\rho^2 = - {{\sin^2 (\pi\tilde\omega)}
\over {(\pi\tilde\omega)^2}} ,
\end{equation}
which is well-known from random matrix theory\cite{mehta}.  Similar 
manipulations lead to identical results when $C(\omega)$ is evaluated 
using the purely Bosonic or mixed expressions analogous to
Eq.~(\ref{CFermion0}).

\subsection{Extension to parametric correlations}
\label{sec:par_corr}

The above calculation can be extended to the case where the random
potential $V_n(x)$ depends on a parameter, $\lambda$:
\begin{eqnarray}
{\cal H}_1(\lambda) &=& \sum_n \int dx \ V_n(x,\lambda) \psi_n^\dagger
\psi_n^{\vphantom{\dagger}} , \\
V_n(x,\lambda) &=& V_n(x) - \lambda W_n(x) .
\end{eqnarray}
The extra term $\lambda W_n(x)$ will be called the ``background
potential''.  For simplicity we take it to be independent of $x$:
$W_n(x) = h_n$.  (It can be shown that the final result 
(\ref{parametric_correlator}) remains essentially unchanged when this 
assumption is lifted.)  We assume that there are no correlations 
between the background potential $\lambda h_n$ and the random potential 
$V_n(x)$.  For the moment the other statistical properties of the 
numbers $h_n$ are left unspecified.

The object of interest is the parametric density of states 
correlation function
\begin{equation}
C(\omega,\lambda) = \left[ \rho(\omega,\lambda) \rho(0,0) \right]_{\rm ens}
- \bar\rho^2 .
\end{equation}
By the same steps as in the previous section, its calculation reduces 
to that of
\[
\sum_{n,n'} \left[ G_{+n}(\omega/2,\lambda/2) G_{-n'}(-\omega/2,-\lambda/2) 
\right]_{{\rm ens},c} .
\]
The calculation bears much similarity to the one we did for $\lambda = 0$,
and we will now focus on the new features arising for $\lambda \not= 0$.

The background potential gives rise to an additional piece in the
Lagrangian,
\begin{equation}
{\cal L}_n \to {\cal L}_n + 
{i \over 2} \lambda h_n \left( \phi_n^* \sigma_z^{\vphantom{*}}
\phi_n^{\vphantom{*}} + \bar\psi_n \sigma_z \psi_n \right) ,
\end{equation}
and in the spin Hamiltonian,
\begin{eqnarray}
        H \to H + \lambda H_{\rm M}, \quad
        H_{\rm M} = i \sum_n h_n ( S_n^z + J_n^z ) ,
\label{Hlambdadef}
\end{eqnarray}
which acts like the coupling to an imaginary magnetic field.
Note that adding $\lambda H_{\rm M}$ is the same as 
substituting $\omega \to \omega + \lambda h_n$.  To prevent the 
background potential from causing an overall shift of the frequency, we 
shall require that $\sum_n h_n$ vanishes identically for every realization 
of the disorder.

What is the effect of $H_{\rm M}$ inside the zero energy multiplet?
By the requirement $\sum_n h_n = 0$, the zero mode approximation
$S_n^z + J_n^z \to N^{-1} \sum_n (S_n^z + J_n^z)$ for $H_{\rm M}$
gives a vanishing result to linear order in $\lambda$.  To see the
effect of the background potential, we must go to higher order and 
do a calculation similar in spirit to that performed by Simons and 
Altshuler\cite{Simons93} for the case of time-reversal invariant disordered
metallic grains.  What the perturbation $H_{\rm M}$ does is to couple the 
zero energy multiplet with the one-magnon sector.  By second-order perturbation
theory, this coupling results in an effective Hamiltonian, $H_\lambda$,
acting on the states of the zero energy multiplet:
\begin{equation}
H_\lambda = H_{\rm M} \left( - \sum_{k\not= 0} {\Pi_k \over D k^2} 
\right) H_{\rm M} ,
\label{effective_Hamiltonian}
\end{equation}
where $\Pi_k$ is the projector onto the one-magnon sector with  momentum $k$.  
In the appendix it is shown that for $N$ large, $H_\lambda$ reduces to
\begin{equation}
H_\lambda = \gamma {\cal C}^{+-} , \quad
\gamma = {\lambda^2 \over N} \sum_{k\not= 0} {|{\tilde h}_k|^2 \over Dk^2} , 
\label{H_gamma}
\end{equation}
where
\begin{equation}
{\cal C}^{+-} = S^+ S^- + J^+ J^- + {\tilde A}^+ A^- - {A}^+ {\tilde A}^- ,
\end{equation}
and 
\begin{equation}
{\tilde h}_k = {1\over N} \sum_n e^{ikn} h_n 
\end{equation}
is the Fourier transform of the background potential.
In order for second-order perturbation theory to apply, the change in 
energy due to the perturbation must be much smaller than the smallest 
energy denominator.  Because the gap for single magnon excitations is of
the order of $D / N^2$, and $H_\lambda$ will be seen below to shift the 
energy by an amount of the order of $\gamma N^2$, the condition
\begin{equation}
\gamma \ll D / N^4
\end{equation}
on $\gamma$ is sufficient for the perturbative formula
(\ref{effective_Hamiltonian}) to be valid.  To give this result some
added physical meaning, consider the simple case in which the numbers
$h_n = \hat{h}_n - \bar{h}$, where the $\hat{h}_n$ are chosen to be
independent, identically distributed random variables with zero mean
and unit variance, and $\bar{h} = N^{-1}\sum_n \hat{h}_n$ is chosen to
maintain zero spatial average.  The ensemble average of the quantity
$|{\tilde h}_k|^2$ is $1/N$ (for $k \neq 0$), and
\begin{equation}
[ \gamma ]_{\rm ens} = {\lambda^2 \over N^2} \sum_{k\not= 0} {1 \over Dk^2}
= {{\lambda^2} \over {12 D}} \label{gamma_ens}
\end{equation}
in the large $N$ limit.  More generally, we are led to define a
dimensionless, order one, $\gamma_0$ by
\begin{equation}
  \gamma =: {{\lambda^2} \over D}\gamma_0. \label{gamma0_def}
\end{equation}
In the particular example here, comparison of
Eqs.~(\ref{gamma_ens}-\ref{gamma0_def}) gives $[\gamma_0]_{\rm ens.} =
1/12$.  The condition on the smallness of $\gamma$ then becomes
\begin{equation}
\lambda \ll D / N^2 ,
\end{equation}
i.e. the typical strength of fluctuation of the background potential
must be much smaller than the one-magnon gap.

We are now ready to compute the parametric density of states correlator,
$C(\omega,\lambda)$. 
{}From Eqs.~(\ref{comrel1}-\ref{comrel4}) and similar commutation relations
involving the lowering operators $S^-$, $J^-$, $A^-$, and ${\tilde A}^-$,
one finds that ${\cal C}^{+-}$ commutes with the Fermionic charges
$Q_\alpha$ and $\overline{Q}_\alpha$, which means that ${\cal C}^{+-}$
is a Casimir invariant for ${\cal H} = {\rm u}(1|1) \oplus {\rm u}(1|1)$.
Therefore ${\cal C}^{+-}$ must be proportional to unity on the
quadruplets:
\begin{equation}
{\cal C}^{+-} | n_1^{\vphantom{*}} n_2^{\vphantom{*}} , \mu\nu \rangle
= \Gamma(n_1 , n_2)| n_1^{\vphantom{*}} n_2^{\vphantom{*}} , \mu\nu \rangle.
\end{equation}
To calculate the eigenvalue $\Gamma(n_1 , n_2)$, we may work out
the action of ${\cal C}^{+-}$ on any one of the four states 
$(\mu,\nu = 0,1)$, say $| n_1^{\vphantom{*}} n_2^{\vphantom{*}} , 01\rangle$.
A straightforward calculation gives
\begin{eqnarray}
\Gamma(n_1 , n_2) &=& n_2 (N+n_2-1) 
\nonumber \\
&+& n_1(N-n_1-1) .
\end{eqnarray}
By including the contribution from $H_\lambda = \gamma {\cal C}^{+-}$
into the Boltzmann weight, we obtain for the parametric density of 
states correlator,
\begin{eqnarray}
C(\omega,\lambda) &=& {\beta^2\over {2\pi^2}} {\rm Re} \sum_{n_2=1}^\infty
\sum_{n_1=0}^{N-1} e^{-i\beta\omega (n_2+n_1)}
\nonumber \\
&&\hspace{2cm} \times e^{-\beta\lambda^2 D^{-1}\gamma_0 \Gamma(n_1,n_2)} .
\label{parametric_correlator}
\end{eqnarray}
For $\beta N \lambda^2 D^{-1} \gamma_0 \ll 1$ or, equivalently, 
$\lambda^2 D^{-1} \gamma_0 \ll \Delta$, we may again take the continuum limit 
and replace the sum by an integral.  By performing a simple but revealing 
substitution of integration variables, we get
\begin{equation}
{C(\omega,\lambda) \over \bar\rho^2} = \int_{|p| \ge \pi} 
\int_{|q| \le \pi} e^{i \tilde\omega (p-q) - \tilde\lambda^2 (p^2 - q^2)} 
{dp dq \over (2\pi)^2} , 
\end{equation}
with the rescaled parameter $\tilde\lambda$ given by
\begin{equation}
\tilde\lambda^2 = \beta {{N^2} \over D} \lambda^2 \gamma_0 / (2\pi)^2 .
\end{equation}
Upon making the identification of $\tilde\omega$ with position and
$\tilde\lambda^2$ with imaginary time, the expression for
$C(\omega,\lambda)$ is seen to
coincide with the dynamical density-density correlation function of a
free Fermi gas.  For the case of small metallic particles, this
coincidence was recently pointed out by Simons et al.\cite{sla}

\section{Conclusions}

\subsection{Summary of results}

We have studied the zero-temperature properties of the
surface states of a three-dimensional layered integer quantum Hall
sample.  Previous work\cite{Chalker,Balents96}\ showed that for weak disorder
in the thermodynamic limit this system is a {\sl dirty chiral metal},
with finite conductivity along the field axis at zero temperature, and
ballistic transport transverse to it.  

In this paper, several new results are obtained and placed in a more
powerful conceptual framework.  By constructing an appropriate
generating functional for disorder averaged correlations, we showed
that this system can in fact be viewed as a kind of supersymmetric (SUSY)
ferromagnetic spin chain.  Like for an ordinary ferromagnet, the ground
state and single magnon excitations can be obtained exactly.  This
allows an exact calculation of the diffuson (or density-density
correlator).  The conventional $z=2$ dynamical scaling for ferromagnetic
spin waves corresponds to the combination of finite conduction
(transverse diffusion) and ballistic transport discussed above.  In
the thermodynamic limit, therefore, we have demonstrated the absence
of localization in the 2d chiral metal, even in the limit of strong
disorder.  

In addition, the quantum formulation of the SUSY ferromagnet provides
a novel method to study the mesoscopics of the chiral metal.  As an
example, we have determined the correlation function of the density of
states at two energies and for two slightly different realizations of
disorder, the so-called parametric correlations.  As expected, in the
ergodic regime this takes on the universal form appropriate for the
unitary ensemble, i.e. a free Fermion density-density correlation
function\cite{Simons93}.  

\subsection{Connection to the non-linear $\sigma$ model}

It is useful to connect the approach taken here to the more
conventional NL$\sigma$M method\cite{efetov}.  This
connection may be made quite explicit using the SUSY coherent states
developed in the appendix.  We will construct a coherent state path
integral representation for the partition function ${\cal Z}$.  We begin by
considering the problem of decoupled edges, i.e. with $D=0$.  Then
\begin{equation}
{\cal Z} = \prod_n {\cal Z}_n,
\end{equation}
where
\begin{equation}
{\cal Z}_n = {\rm STr} e^{-\beta H_n}.
\end{equation}
The single super spin Hamiltonian for $\lambda = 0$ is determined by 
the condition $H_\eta = \sum_n H_n$, so by Eq.~(\ref{Hetadef}),
        \begin{equation}
        H_n = 2\eta \left (S_n^z + J_n^z \right) .
        \end{equation}
The coherent state path integral can then be simply obtained using the 
resolution of unity, Eq.~(\ref{identity}), and the formula for the 
supertrace, Eq.~(\ref{supertrace}).  One finds 
\begin{equation}
{\cal Z}_n = \int D(Z_n,\tilde Z_n) e^{-S_n[Z_n]},
\end{equation}
where the single mode action is
        \[
        S_n[Z_n] = \int_0^\beta d\tau \ \langle Z_n | \partial_\tau
        + H_n | Z_n \rangle \equiv \int_0^\beta d\tau {\cal L}_n.
        \]
The single spin Lagrangian is given by
        \begin{eqnarray}
        {\cal L}_n &=& \langle Z_n | \partial_\tau + H | Z_n \rangle
        \nonumber \\
        &=& -{\rm STr} (1+\tilde Z_n Z_n)^{-1} \tilde Z_n 
        (\partial_\tau + 2\eta) Z_n .
        \label{single_spin_Lagrangian_Z}
        \end{eqnarray}
This expression can be put into a simpler form by defining the $4\times 4$ 
supermatrices
\begin{equation}
g_n = \pmatrix{1 &Z_n\cr -\tilde Z_n &1\cr}, \qquad \Lambda =
\pmatrix{1 &0\cr 0 &-1\cr}
\end{equation}
and
\begin{equation}
Q_n = g_n^{\vphantom{-1}} \Lambda g_n^{-1}.
\end{equation}
In terms of these fields, the single spin Lagrangian becomes
        \begin{equation}
        {\cal L}_n = {1\over 2} {\rm STr} \left( \Lambda g_n^{-1}
        \partial_\tau^{\vphantom{-1}} g_n^{\vphantom{-1}} \right)
        + {\eta\over 2} {\rm STr}\left(\Lambda Q_n\right) .
        \label{single_spin_Lagrangian}
        \end{equation}
The first (dynamical phase) term is of the Wess-Zumino type, and {\sl
cannot} be written in a globally non-singular form in terms of the $Q$
field.  A Wess-Zumino term also occurs in the coherent state path
integral for an ordinary SU(2) spin\cite{Fradkin}, and could have been
expected here on general grounds.  It is necessary to obtain the $z=2$
dynamics appropriate for a ferromagnet.  Inclusion of such a term
allows us here to obtain a true NL$\sigma$M formulation, in contrast
to the earlier $Q$-matrix formulation in Ref.\cite{Balents96},
in which only an expansion in the ordered (metallic) phase was
determined.

Eq.~(\ref{single_spin_Lagrangian}) holds for a single super spin.  It
is now straightforward to include the exchange coupling $H_D$ to
obtain 
\begin{equation}
{\cal Z} = \int D(Z,\tilde Z) e^{-S}
\end{equation}
where
\begin{equation}
S = \int_0^\beta d\tau \sum_n \left( {\cal L}_n + {\cal L}_{{\rm
int},n} \right),
\label{fullNLSM}
\end{equation}
and
\begin{equation}
{\cal L}_{{\rm int},n} = {D\over 4} {\rm STr} \left( Q_n Q_{n+1} \right).
\end{equation}
Eq.~(\ref{fullNLSM}) is the full NL$\sigma$M action for the 2d system.
The universal level correlations, usually discussed in this context,
are properties of the zero-dimensional quantum limit, in which it is
appropriate to make a ``zero mode'' approximation, neglecting spatial
and temporal (i.e. $\tau \equiv x$) variations of $Q$.  The zero mode
theory then becomes a single super-integral over $Q_0$.   This
formulation can be connected back to the operator one of the text, by
noting that the resulting integral is simply another representation of
the restricted trace over the zero energy multiplet of completely polarized
super spin states.

\subsection{Questions and open problems}

We conclude with a discussion of various questions which remain open
to future investigation.  One interesting issue is to understand in
more detail the nature of the crossovers between the three mesoscopic
regimes described in Section V.  In the language of the ferromagnetic
super spin chain, which has finite length $N$ and is at finite
temperatures ($\beta < \infty$), the three regimes correspond to: (i)
a ``zero-dimensional" regime in which all (finite wavevector)
spin-waves are absent, being too costly in energy, (ii) a ``1d
diffusive" regime in which the spins behave classically but are still
ferromagnetically ordered, and (iii) a ``localized" regime in which
the spin chain length exceeds the ferromagnetic spin correlation
length.  For ordinary (non-super) ferromagnetic spin chains, recent
progress has been made in computing various scaling functions
connecting these regimes\cite{Takahashi96}. The apparent generality of
the technique suggests that progress might likewise be made for the
super spin chain.\cite{subirnick}

Also of interest in terms of crossover behavior are the scaling
properties of the {\sl wavefunctions} in various limits.  Although we
do not expect interesting scaling properties in the thermodynamic
limit (because the 2d system is truly a stable (chiral) metal),
interesting possibilities arise in the intermediate mesoscopic regime.
Specifically, in the ``1d diffusive" regime, with $L_0 = \sqrt{D\beta}
\ll N \ll \xi=D\beta$, the electron motion is ergodic on frequency
scales $\omega \sim \Delta$.  Nevertheless, there is a breakdown of
the zero-mode approximation, which requires $L_0 \gg N$.  We then
expect the wavefunctions to become very broadly distributed random
variables\cite{Falko-Efetov}. It would be interesting to see if the
moments exhibit multifractal scaling in this regime, as well as to
look at scaling of multi-point correlators.

Thirdly, it would be interesting to calculate the conductance 
fluctuations\cite{Fluctuations},
which would involve applying appropriate boundary conditions to the
ends of the super spin chain and calculating an eight-Fermion (or
eight-Boson, etc.) correlator (product of four Green's functions).
Preliminary work using diagrammatic techniques by Mathur\cite{Mathur}\
suggests rather interesting crossover phenomena in the variance of
the conductance.

Lastly, these methods may be useful to connect with earlier work on
chiral classical wave propagation in a disordered medium.  As pointed
out in Ref.~\cite{Saul92}, the Schr\"odinger equation for
electrons in the surface sheath may be viewed as a classical wave
equation for ``directed'' propagation.  It is then of interest to
study the spreading and deflection of a point source at an initial
$x=0, n=0$ to some larger $x$.  The beam width is closely related to
our diffuson, but (like the conductance fluctuations) the motion of
the beam center involves the product of four Green's functions, but
may also be tractable\cite{Mathur}.

\noindent {\sl Acknowledgements}

We are grateful to Nick Read, Subir Sachdev and particularly Harsh
Mathur for illuminating conversations.  This work has been supported
by the National Science Foundation under grants No. PHY94-07194,
DMR-9400142 and DMR-9528578.  M.R.Z. acknowledges partial support from
the Deutsche Forschungsgemeinschaft by SFB 341,
K\"oln-Aachen-J\"ulich.

\section*{Appendix}

This appendix is in two parts.  In the first part we are going to
develop some of the mathematical structures underlying the coherent
state path integral for the super spin Hamiltonian.  In the second
part we will derive the expression (\ref{H_gamma}) for the effective
zero mode Hamiltonian $H_\lambda$, which results from treating the
background potential by second-order perturbation theory.

Our first item will be to discuss the group of canonical transformations
of the Bose and Fermi operators.  For that purpose we introduce a more
economical notation, by setting
\begin{eqnarray}
        c_{B\uparrow} &=& B_{\uparrow}, \quad 
        c_{F\uparrow} = F_{\uparrow}, \quad 
        c_{B\downarrow} = B_{\downarrow}, \quad 
        c_{F\downarrow} = F_{\downarrow},
        \nonumber \\
        \overline{c}_{B\uparrow} &=& \overline{B}_{\uparrow}, \quad 
        \overline{c}_{F\uparrow} = F_{\uparrow}^\dagger, \quad 
        \overline{c}_{B\downarrow} = \overline{B}_{\downarrow}, \quad 
        \overline{c}_{F\downarrow} = F_{\downarrow}^\dagger .
        \nonumber
        \end{eqnarray}
Now let $A$, $B$, $C$ and $D$ be complex $2\times 2$ supermatrices, i.e.
        \[
        A = \pmatrix{A_{FF} &A_{FB}\cr A_{BF} &A_{BB}\cr}
        \quad {\rm etc.} ,
        \]
where $A_{FF}$ and $A_{BB}$ are complex numbers, while $A_{FB}$ and
$A_{BF}$ are Grassmann numbers.  Consider then the transformation
        \begin{eqnarray}
        &&\overline{c}_{a\uparrow}^n \to \overline{c}_{b\uparrow}^n 
        A_{ba}^{\vphantom{n}} + \overline{c}_{b\downarrow}^n 
        C_{ba}^{\vphantom{n}} ,
        \nonumber \\
        &&\overline{c}_{a\downarrow}^n \to \overline{c}_{b\uparrow}^n 
        B_{ba}^{\vphantom{n}} + \overline{c}_{b\downarrow}^n 
        D_{ba}^{\vphantom{n}} ,
        \nonumber
        \end{eqnarray}  
where the superscript $n = 1, ... , N$ indexes the chiral modes and, 
here and below, the summation convention is used.  By setting 
$\{ \overline{c}_\alpha \} = \{ \overline{c}_{F\uparrow},
\overline{c}_{B\uparrow}, \overline{c}_{F\downarrow} ,
\overline{c}_{B\downarrow} \}$, we write this in the abbreviated form
        \[
        \overline{c}_{\alpha}^n \to \overline{c}_{\beta}^n 
        g_{\beta\alpha}^{\vphantom{n}}, \qquad g = \pmatrix{A &B\cr C &D\cr} .
        \]
If $g$ has an inverse, we can also transform the annihilators, by
        \[
        c_\alpha^n \to (g^{-1})_{\alpha\beta}^{\vphantom{n}} c_\beta^n .
        \]
When rearranging products, we follow the convention that Grassmann
numbers not only anticommute among themselves, but also anticommute
with the Fermionic operators.  It is then easy to see that the
transformation $c_\alpha^n \to (g^{-1})_{\alpha\beta}^{\vphantom{n}}
c_\beta^{\vphantom{n}}$, $\overline{c}_{\alpha}^n \to
\overline{c}_{\beta}^n g_{\beta\alpha}^{\vphantom{n}}$, is canonical, 
i.e. preserves the graded commutator.

Canonical transformations have an inverse, and a succession of
two canonical transformations is again canonical.  Therefore such
transformations\cite{bb} form a group, which in the present context is 
the Lie supergroup ${\rm Gl}(2|2)$, obtained by complexifying
${\rm U}(2|1,1)$.  The elements $g$ of ${\rm Gl}(2|2)$ can be written 
in the form of a Gauss decomposition,
        \[
        g = \pmatrix{A &B\cr C &D\cr} = 
        \pmatrix{1 &Z\cr 0 &1\cr} \pmatrix{h_\uparrow &0\cr 
        0 &h_\downarrow\cr} \pmatrix{1 &0\cr \tilde Z &1\cr} ,
        \]
where, from an easy calculation,
        \begin{eqnarray}
        Z &=& B D^{-1} , \quad h_\downarrow = D ,
        \nonumber \\
        \tilde Z &=& D^{-1} C , \quad h_\uparrow = A - B D^{-1} C .
        \nonumber 
        \end{eqnarray}

Every canonical transformation $g$ can be realized by a Fock space
operator $T_g$ that is obtained by exponentiating some bilinear
$\overline{c} c$, and acts by ${c}_\alpha^n \to T_g^{\vphantom{-1}}
{c}_\alpha^n T_g^{-1}$, $\overline{c}_\alpha^n \to T_g^{\vphantom{-1}}
\overline{c}_\alpha^n T_g^{-1}$.  In detail the correspondences are as
follows:
        \begin{eqnarray}
        T_{\pmatrix{1 &Z\cr 0 &1\cr}} &=&  \exp \left( 
        \overline{c}_{a\uparrow}^n Z_{ab}^{\vphantom{n}} 
        c_{b\downarrow}^n \right) , 
        \nonumber \\
        T_{\pmatrix{1 &0\cr \tilde Z &1\cr}} &=&        
        \exp \left( \overline{c}_{a\downarrow}^n 
        {\tilde Z}_{ab}^{\vphantom{n}} c_{b\uparrow}^n \right) ,
        \nonumber \\
        T_{\pmatrix{e^A &0\cr 0 &1\cr}} &=&
        \exp \left( \overline{c}_{a\uparrow}^n A_{ab}^{\vphantom{n}} 
        c_{b\uparrow}^n \right) ,
        \nonumber \\
        T_{\pmatrix{1 &0\cr 0 &e^D\cr}} &=&
        \exp \left( \overline{c}_{a\downarrow}^n D_{ab}^{\vphantom{n}} 
        c_{b\downarrow}^n \right) .
        \nonumber
        \end{eqnarray}  
By construction, the correspondence $g \to T_g$ defines a
representation of ${\rm Gl}(2|2)$ on Fock space. Therefore every
relation that is valid for supermatrices $g$, and uses no more than
the Lie supergroup structure of ${\rm Gl}(2|2)$, also holds for the
Fock space operators $T_g$.  In particular, by applying the Gauss
decomposition to the product
        \[
        \pmatrix{1 &0\cr \tilde Z &1\cr} \pmatrix{1 &Z\cr 0 &1\cr} ,
        \]
we get
        \begin{eqnarray}
        &&\exp \left( \overline{c}_{a \downarrow}^n \tilde 
        Z_{ab}^{\vphantom{n}} c_{b \uparrow}^n \right)
        \exp \left( \overline{c}_{a \uparrow}^n Z_{ab}^{\vphantom{i}} 
        c_{b \downarrow}^n \right)
        \nonumber \\
        &=& \exp \left( \overline{c}_{a\uparrow}^n 
        [Z(1+\tilde Z Z)^{-1}]_{ab}^{\vphantom{n}} 
        c_{b \downarrow}^n \right)
        \nonumber \\
        &&\times \exp \left( \overline{c}_{a\downarrow}^n 
        [\ln (1+\tilde Z Z)]_{ab}^{\vphantom{n}} c_{b\downarrow}^n 
        - \overline{c}_{a\uparrow}^n [\ln (1+Z\tilde Z)]_{ab}^{\vphantom{n}} 
        c_{b\uparrow}^n \right)
        \nonumber \\
        &&\times \exp \left( \overline{c}_{a\downarrow}^n 
        [\tilde Z (1+Z\tilde Z)^{-1}]_{ab}^{\vphantom{n}} 
        c_{b\uparrow}^n \right) , \nonumber
        \end{eqnarray}
which is called a ``disentangling'' identity, since it moves the
${\rm Gl}(2|2)$ lowering operators all the way to the right, and
the raising operators all the way to the left.

Consider now the generalized coherent states\cite{perelomov}
        \begin{eqnarray}
        | Z \rangle &=&
        \exp \left( \overline{c}_{a\uparrow}^n Z_{ab}^{\vphantom{i}} 
        c_{b\downarrow}^n \right) | 0 \rangle \ {\rm SDet}(1+\tilde Z Z)^{N/2},
        \nonumber \\
        \langle Z | &=& {\rm SDet}(1+\tilde Z Z)^{N/2} \langle 0 |
        \exp \left( \overline{c}_{b\downarrow}^n 
        \tilde Z_{ba}^{\vphantom{n}} c_{a\uparrow}^n \right) .
        \nonumber
        \end{eqnarray}
The superdeterminant of a $2\times 2$ supermatrix,
        \begin{eqnarray}
        &&{\rm SDet}\pmatrix{M_{FF} &M_{FB}\cr M_{BF} &M_{BB}\cr}
        \nonumber \\
        &=& \left( M_{BB} - M_{BF} {M_{FF}}^{-1} M_{FB} \right) / M_{FF} 
        \nonumber \\
        &=& M_{BB} / \left( M_{FF} - M_{FB} {M_{BB}}^{-1} M_{BF} \right) ,
        \nonumber
        \end{eqnarray}
satisfies $\ln {\rm SDet} M = {\rm STr} \ln M$, where ${\rm STr}M
= - M_{FF} + M_{BB} =: \sum_a (-1)^{|a|} M_{aa}$, with $|a| = 0$ if
$a = B$ (Boson) and $|a| = 1$ if $a = F$ (Fermion), is the supertrace.
Using this, the disentangling identity, and the relations
        \[
        \overline{c}_{a\uparrow}^n {c}_{b\uparrow}^n | 0 \rangle = 0 , 
        \qquad
        \overline{c}_{a\downarrow}^n {c}_{b\downarrow}^n | 0 \rangle 
        = \delta_{ab} (-1)^{|a|+1} N | 0 \rangle ,
        \]
it is easy to check that the generalized coherent states are normalized,
        \[
        \langle Z | Z \rangle = 1 .
        \]
Another useful way of writing the generalized coherent states is
        \[
        | Z \rangle = T_g | 0 \rangle , \qquad
        \langle Z | = \langle 0 | T_g^{-1} ,
        \]
where
        \begin{eqnarray}
        g &=& \pmatrix{1 &Z\cr 0 &1\cr} \pmatrix{ (1+Z \tilde Z)^{+1/2} &0\cr
        0 &(1+\tilde Z Z)^{-1/2} \cr} \pmatrix{1 &0\cr -\tilde Z &1\cr}
        \nonumber \\
        &=& \pmatrix{ (1+Z\tilde Z)^{-1/2} &Z(1+{\tilde Z}Z)^{-1/2} \cr
        -{\tilde Z}(1+Z{\tilde Z})^{-1/2} &(1+{\tilde Z}Z)^{-1/2} \cr} ,
        \label{from_g_to_Z}
        \end{eqnarray} 
and
        \begin{equation}
        g^{-1} = \pmatrix{ (1+Z\tilde Z)^{-1/2} &-Z(1+{\tilde Z}Z)^{-1/2} \cr
        {\tilde Z}(1+Z{\tilde Z})^{-1/2} &(1+{\tilde Z}Z)^{-1/2} \cr} .
        \label{from_g_minus_one_to_Z}
        \end{equation}

The zero energy multiplet of Sec.~\ref{sec:zero_energy} coincides
with the space of states obtained by acting repeatedly with the
${\rm Gl}(2|2)$ raising operators ${\bar c}_{a\uparrow}^n c_{b\downarrow}^n$
on the vacuum $| 0 \rangle$. In\cite{mrz-circular} it was shown that,
if $D(Z,\tilde Z) = Dg_H^{\vphantom{+}}$ is the uniform superintegration
measure on the
supermanifold $G / H = {\rm Gl}(2|2) / {\rm Gl}(1|1)\times{\rm Gl}(1|1)$
parameterized by the coherent states $| Z \rangle = T_g | 0 \rangle$,
the unit operator on the zero energy multiplet can be resolved by
        \begin{equation}
        1 = \int Dg_H^{\vphantom{+}} T_g^{\vphantom{-1}} | 0 \rangle 
        \langle 0 | T_g^{-1} = \int D(Z,\tilde Z) | Z \rangle \langle Z | .
        \label{identity}
        \end{equation}
(Note that the conventions of that reference differ from ours by
$\tilde Z \to - \tilde Z$ and the exchange of the Fermionic and
Bosonic sectors.) The integral is over
        \[
        {\tilde Z}_{FF} = {\bar Z}_{FF} , \qquad
        {\tilde Z}_{BB} = - {\bar Z}_{BB} ,
        \]
with the integration domain being
        \[
        0 \le |Z_{FF}|^2 < \infty , \qquad
        0 \le |Z_{BB}|^2 < 1 .
        \]
These relations mean\cite{mrz-iqhe} that the variable $Z_{FF}$ is a 
complex stereographic coordinate for the two-sphere ${\rm S}^2$ (as is
well-known for SU(2) coherent states) while $Z_{BB}$ parameterizes a
two-hyperboloid ${\rm H}^2$ (as is appropriate for the non-compact group 
SU(1,1)).  Of course $Z_{BF}$, $Z_{FB}$, ${\tilde Z}_{BF}$, and 
${\tilde Z}_{FB}$ are Grassmann variables.

Given the resolution of unity (\ref{identity}), we can convert
supertraces over the zero energy multiplet into integrals over the
generalized coherent states:
        \begin{equation}
        \langle O \rangle_0 = {\rm Tr}_0 (-1)^{N_F} O =
        \int D(Z,\tilde Z) \langle Z | O | Z \rangle .
        \label{supertrace}
        \end{equation}
The coherent state path integral for the partition function of
a super spin system with Hamiltonian $H$, is now obtained in the
usual manner by inserting resolutions of unity between 
infinitesimal imaginary time slices of the Boltzmann weight
$\exp (-\beta H)$.  The Lagrangian of the resulting path integral
is ${\cal L} = \langle Z | \partial_\tau + H | Z \rangle$, whose
explicit form is easily calculated by using the definition of the 
coherent states and the disentangling identity.  For example,
        \begin{eqnarray}
        &&\langle Z | \partial_\tau | Z \rangle - \partial_\tau
        \ln {\rm SDet}(1+{\tilde Z}Z)^{N/2}
        \nonumber \\
        &=& {\rm SDet}(1+{\tilde Z}Z)^N \langle 0 |
        \exp ( {\bar c}_{b\downarrow}^n {\tilde Z}_{ba}^{\vphantom{n}}
        c_{a\uparrow}^n) \partial_\tau
        \exp ( {\bar c}_{a\uparrow}^n Z_{ab}^{\vphantom{n}} 
        c_{b\downarrow}^n) | 0 \rangle
        \nonumber \\
        &=& {\rm SDet}(1+{\tilde Z}Z)^N {\partial\over\partial s}\Big|_{s=0}
        {\rm SDet}(1+{\tilde Z}Z+s{\tilde Z}{\partial_\tau Z})^{-N}
        \nonumber \\
        &=& - N {\rm STr} \left[ 
        (1+{\tilde Z}Z)^{-1}{\tilde Z}{\partial_\tau Z} \right] .
        \nonumber
        \end{eqnarray}
The Wess-Zumino (or linear in $\partial_\tau$) term in 
(\ref{single_spin_Lagrangian_Z}) is obtained on setting the number 
of chiral modes $N$ equal to one and omitting a total $\tau$ derivative.

Below we will need an explicit expression for the coherent state
expectation value of the operator ${\cal C}^{+-}$ appearing in the
formula (\ref{H_gamma}) of $H_\lambda$.  To calculate it we write
        \begin{eqnarray}
        {\cal C}^{+-} 
        &=& S^+ S^- + J^+ J^- + {\tilde A}^+ A^- - A^+ {\tilde A}^-
        \nonumber \\
        &=& S^- S^+ + J^- J^+ + {\tilde A}^- A^+ - A^- {\tilde A}^+ .
        \nonumber
        \end{eqnarray}
The second equality is invalid for both ${\rm SU}(2)$ and 
${\rm SU}(1,1)$, since $S^+ S^- \not= S^- S^+$ and $J^+ J^- \not=
J^- J^+$, but it does hold in the present case as a result of 
cancellations due to supersymmetry.  Using the second form of 
${\cal C}^{+-}$ we easily find
        \begin{eqnarray}
        &&{\rm SDet}(1+{\tilde Z}Z)^{-N} \langle Z | {\cal C}^{+-} | Z \rangle
        \nonumber \\
        &=& \sum (-1)^{|d|+1} {\partial^2 \over \partial Z_{cd} 
        \partial {\tilde Z}_{dc}} \
        \langle 0 | 
        \exp ( {\bar c}_{b\downarrow}^n {\tilde Z}_{ba}^{\vphantom{n}}
        c_{a\uparrow}^n) \exp ( {\bar c}_{a\uparrow}^n Z_{ab}^{\vphantom{n}} 
        c_{b\downarrow}^n) | 0 \rangle
        \nonumber \\
        &=& \sum (-1)^{|b|+1} {\partial^2 \over \partial Z_{ab} 
        \partial {\tilde Z}_{ba}} \ {\rm SDet}(1+{\tilde Z}Z)^{-N} ,
        \nonumber
        \end{eqnarray}
which yields
        \begin{equation}
        \langle Z | {\cal C}^{+-} | Z \rangle = 
        - N^2 {\rm STr} \left[ {\tilde Z}Z (1+{\tilde Z}Z)^{-2} \right]
        + {\cal O}(N^1) .
        \label{coh_state_exp_value_C}
        \end{equation}

We now turn to the effective zero mode Hamiltonian
        \[
        H_\lambda = H_{\rm M} \left( - \sum_{k\not= 0} {\Pi_k \over D k^2} 
        \right) H_{\rm M} ,
        \]
and show that it reduces to (\ref{H_gamma}) in the large-$N$ limit.  The
first step is to construct the projector on the space of momentum-$k$ 
single magnon excitations, $\Pi_k$.  Magnons can be created on top of 
the vacuum or any other state of the zero energy multiplet.  We define 
the single magnon creation and annihilation operators by
        \begin{eqnarray}
        {\cal J}_{ab}^+(k) &=& \sum_n e^{+ikn} \ \overline{c}_{a\uparrow}^n
        c_{b\downarrow}^n ,
        \nonumber \\
        {\cal J}_{ab}^-(k) &=& \sum_n e^{-ikn} \ \overline{c}_{a\downarrow}^n
        c_{b\uparrow}^n ,
        \nonumber
        \end{eqnarray}
and consider the states
        \[
        | ab , k \rangle = {1\over\sqrt{N}} 
        {\cal J}_{ab}^{+}(k) | 0 \rangle .
        \]
The corresponding projection operator is
        \[
        \pi_k = \sum_{ab} | ab , k \rangle \langle ab , k | 
        = {1 \over N} \sum_{ab} (-1)^{|b|+1} 
        {\cal J}_{ab}^+(k) | 0 \rangle \langle 0 | {\cal J}_{ba}^- (k) .
        \]
It is easy to verify that, with the normalization and sign factors
chosen, $\pi_k$ satisfies $\pi_k \pi_k = \pi_k$, and is a singlet
with respect to the (complexified) algebra ${\cal H} = {\rm gl}(1|1) 
\oplus {\rm gl}(1|1)$
generated by the zero mode operators $Q_\uparrow$, $Q_\downarrow$, 
$\overline{Q}_\uparrow$, $\overline{Q}_\downarrow$, $n_{f\uparrow}$, 
$n_{f\downarrow}$, $n_{b\uparrow}$, and $n_{b\downarrow}$. 
The states $| ab , k \rangle$ are single magnon excitations built
on the vacuum.  Other one-magnon states (not based on the vacuum)
are generated by applying a global rotation,
        \[
        | ab , k \rangle \to T_g^{\vphantom{+}} | ab , k \rangle .
        \]
The projector $\pi_k$ transforms as
        \[
        \pi_k \to T_g^{\vphantom{-1}} \pi_k^{\vphantom{-1}} T_g^{-1} .
        \]
The projector on the entire one-magnon sector is obtained by averaging
over all global rotations,
        \begin{equation}
        \Pi_k = \int Dg_H^{\vphantom{+}}
        \ T_g^{\vphantom{-1}} \pi_k^{\vphantom{-1}} T_g^{-1} .
        \label{one_magnon_projector}
        \end{equation}
The integrand is well-defined as a function on the coset space
$G / H$ since, by the ${\cal H}$-singlet property of $\pi_k$,
        \[
        T_{gh}^{\vphantom{-1}} \pi_k^{\vphantom{-1}} T_{gh}^{-1}
        = T_{g}^{\vphantom{-1}} T_h^{\vphantom{-1}} \pi_k^{\vphantom{-1}}
        T_h^{-1} T_{g}^{-1}
        = T_{g}^{\vphantom{-1}} \pi_k^{\vphantom{-1}} T_{g}^{-1}
        \]
for $h \in H = {\rm Gl}(1|1)\times{\rm Gl}(1|1)$.

To go further, we need to take advantage of the large-$N$ limit.
Let us evaluate the expectation value of ${H}_{\rm M} \Pi_k
{H}_{\rm M}$  in a coherent state $| Z \rangle \equiv T_g | 0 \rangle$,
        \[
        E := \int Dg'_H \ \langle 0 | T_g^{-1} H_{\rm M}^{\vphantom{-1}} 
        T_{g'}^{\vphantom{-1}} \pi_k^{\vphantom{-1}} 
        T_{g'}^{-1} H_{\rm M}^{\vphantom{-1}} 
        T_g^{\vphantom{-1}} | 0 \rangle .
        \]
For large $N$ we expect the dominant contributions to the $g'$ integral
to come from the immediate vicinity of $g = g(Z,\tilde Z)$, so
we substitute variables $g' = g \exp X$ (with $X$ subject to the
transversality condition $\Lambda X + X \Lambda = 0$)
and approximate $Dg'_H$ by its linearization $DX$ at $g$.  The
integration over $X$ is trivial since
        \[
        \langle 0 | T_g^{-1} T_{g \exp X}^{\vphantom{-1}} | 0 \rangle
        = \langle 0 | T_{\exp X} | 0 \rangle \simeq
        e^{- N {\rm STr} X^2 / 4} \to \delta(X)
        \]
reduces to a $\delta$-function in the large-$N$ limit.  Thus, doing 
the $g'$ integral to leading order in the large parameter $N$, we get
        \[
        E = \langle 0 | T_g^{-1} H_{\rm M}^{\vphantom{-1}}
        T_g^{\vphantom{-1}} \pi_k^{\vphantom{-1}} T_g^{-1} 
        H_{\rm M}^{\vphantom{-1}} T_g^{\vphantom{-1}} | 0 \rangle .
        \]
The large-$N$ limit is semiclassical in nature, and it is not
hard to see that all off-diagonal matrix elements of $H_\lambda$ 
become negligible for $N\to\infty$.  Therefore, it is in fact sufficient to 
consider the diagonal ones, which is what we are doing.

The next step is to compute the rotated perturbation 
$T_g^{-1} H_{\rm M}^{\vphantom{-1}} T_g^{\vphantom{-1}}$ from the
expression $H_{\rm M} = i \sum_{n,a} h_n ( \overline{c}_{a\uparrow}^n
c_{a\uparrow}^n - \overline{c}_{a\downarrow}^n c_{a\downarrow}^n ) / 2$. 
For this we set
        \[
        g = \pmatrix{A &B\cr C &D\cr}, \quad  
        g^{-1} = \pmatrix{{\tilde A} &{\tilde B}\cr {\tilde C} 
        &{\tilde D}\cr},
        \]
and use the transformation laws (see the beginning of this appendix)
        \begin{eqnarray}
        T_g^{-1} \overline{c}_{a\uparrow}^n T_g^{\vphantom{-1}} 
        &=& \overline{c}_{a'\uparrow}^n {\tilde A}_{a'a}^{\vphantom{n}} 
        + \overline{c}_{a'\downarrow}^n {\tilde C}_{a'a}^{\vphantom{n}} ,
        \nonumber \\
        T_g^{-1} \overline{c}_{b\downarrow}^n T_g^{\vphantom{-1}} 
        &=& \overline{c}_{b'\uparrow}^n {\tilde B}_{b'b}^{\vphantom{n}} 
        + \overline{c}_{b'\downarrow}^n {\tilde D}_{b'b}^{\vphantom{n}} ,
        \nonumber \\
        T_g^{-1} c_{a\uparrow}^n T_g^{\vphantom{-1}}
        &=& {A}_{aa'}^{\vphantom{n}} c_{a'\uparrow}^n 
        + {B}_{aa'}^{\vphantom{n}} c_{a'\downarrow}^n ,
        \nonumber \\
        T_g^{-1} c_{b\downarrow}^n T_g^{\vphantom{-1}}
        &=& {C}_{bb'}^{\vphantom{n}} c_{b'\uparrow}^n 
        + {D}_{bb'}^{\vphantom{n}} c_{b'\downarrow}^n ,
        \nonumber
        \end{eqnarray}
which yield
        \begin{eqnarray}
        T_g^{-1} H_{\rm M}^{\vphantom{-1}} T_g^{\vphantom{-1}}
        &=& {i \over 2} \sum_n h_n \Big( \overline{c}_\uparrow^n 
        (\tilde A B - \tilde B D) c_\downarrow^n 
        \nonumber \\
        &&\hspace{1.5cm}
        + \overline{c}_\downarrow^n (\tilde C A 
        - \tilde D C) c_\uparrow^n + ... \Big) .
        \nonumber
        \end{eqnarray}
On taking the matrix element with the one-magnon creation operator,
we obtain
        \[
        \langle 0 | T_g^{-1} H_{\rm M}^{\vphantom{-1}}
        T_g^{\vphantom{-1}} {\cal J}_{ab}^+(k) | 0 \rangle =
        {iN \over 2} \tilde h_k (-1)^{|a||b|+1} (\tilde C A - 
        \tilde D C)_{ba} .
        \]
Similarly,
        \[
        \langle 0 | {\cal J}_{ba}^-(k) T_g^{-1} H_{\rm M}^{\vphantom{-1}}
        T_g^{\vphantom{-1}} | 0 \rangle = {iN \over 2} \tilde h_{-k} 
        (-1)^{|a||b|+1} (\tilde A B - \tilde B D)_{ab} .
        \]
We now switch notation from $g$ to $Z,\tilde Z$.  {}From 
(\ref{from_g_to_Z},\ref{from_g_minus_one_to_Z}) we read off the relations
        \begin{eqnarray}
        \tilde C A - \tilde D C &=& 2 \tilde Z (1+Z\tilde Z)^{-1} ,
        \nonumber \\
        \tilde A B - \tilde B D &=& (1 + Z\tilde Z)^{-1} 2Z ,
        \nonumber 
        \end{eqnarray}
whence the coherent state expectation value of $H_\lambda$ takes the form
        \[
        \langle Z | H_\lambda | Z \rangle = - N \sum_{k\not =0}
        {\tilde h_k \tilde h_{-k} \over Dk^2} \times 
        {\rm STr} \left[ Z\tilde Z(1+Z\tilde Z)^{-2} \right] .
        \]
Comparison with (\ref{coh_state_exp_value_C}) gives 
$H_\lambda = \gamma {\cal C}^{+-}$, as claimed in (\ref{H_gamma}).

\end{document}